\documentclass[journal,final,twocolumn,10pt,twoside]{IEEEtranTCOM}

\normalsize

\usepackage{multicol,stfloats,graphicx,epsfig,epstopdf,amsmath,amssymb,enumerate,float,color,cite,subfigure}

\ifCLASSINFOpdf
\else
\fi

\begin{document}
%
\title{A WINNER+ Based 3-D Non-Stationary Wideband MIMO Channel Model }
%
%
%

\author{Ji Bian, Jian Sun,~\IEEEmembership{Member,~IEEE}, Cheng-Xiang Wang,~\IEEEmembership{Fellow,~IEEE}, Rui Feng, Jie Huang, Yang~Yang,~\IEEEmembership{Fellow,~IEEE}, and Minggao Zhang

\thanks{Manuscript received December 29, 2016; revised June 23, 2017 and October 9, 2017; accepted December 5, 2017. The authors gratefully acknowledge the support from the the Natural Science Foundation of China (No. 61771293, 61571004, 61371110, 61210002), the Science and Technology Commission of Shanghai Municipality (STCSM) (No. 16ZR1435200), the 863 Project in 5G (No. 2014AA01A707), the EU FP7 QUICK project (No. PIRSES-GA-2013-612652), the EU H2020 5G Wireless project (No. 641985), the EU H2020 RISE TESTBED project (No. 734325), and the EPSRC TOUCAN project (No. EP/L020009/1). The associate editor coordinating the review of this paper and approving it for publication was L. Liu.}

\thanks{J. Bian, J. Sun, R. Feng, J. Huang, and M. Zhang are with Shandong Provincial Key Lab of Wireless Communication Technologies, School of Information Science and Engineering, Shandong Unviersity, Jinan, Shandong, 250100, China (e-mail: bianjimail@163.com, sunjian@sdu.edu.cn, fengxiurui604@163.com, 1101952624@qq.com, zmg225@163.com).}

\thanks{ C.-X. Wang (corresponding author) is with the Institute of Sensors, Signals and Systems, School of Engineering and Physical Sciences, Heriot-Watt University, Edinburgh, EH14 4AS, U.K. He is also with Shandong Provincial Key Lab of Wireless Communication Technologies, School of Information Science and Engineering, Shandong Unviersity, Jinan, Shandong, 250100, China (e-mail: cheng-xiang.wang@hw.ac.uk).}

\thanks{Y. Yang is with Key Lab of Wireless Sensor Network and Communication and Shanghai Research Center for Wireless Communications, SIMIT, Chinese Academy of Sciences, China (e-mail: yang.yang@wico.sh).}

}

%
%

\markboth{IEEE Transactions on Wireless Communications, vol. xx, no. xx, MONTH 2017}%
{}
%



\maketitle
\vspace{-2cm}
\begin{abstract}
In this paper, a three-dimensional (3-D) non-stationary wideband multiple-input multiple-output (MIMO) channel model based on the WINNER+ channel model is proposed.  The angular distributions of clusters in both the horizontal and vertical planes are  jointly considered. The receiver and clusters can be moving, which makes the model more general. Parameters including number of clusters,  powers, delays, azimuth angles of departure (AAoDs), azimuth angles of arrival  (AAoAs), elevation angles of departure (EAoDs), and elevation angles of arrival (EAoAs) are time-variant. The cluster time evolution is modeled using a birth-death process. Statistical properties, including spatial cross-correlation function (CCF), temporal autocorrelation function (ACF), Doppler power spectrum density (PSD), level-crossing rate (LCR), average fading duration (AFD), and stationary interval are investigated and analyzed. The LCR,  AFD, and stationary interval of the proposed channel model are validated against the measurement data. Numerical and simulation results show that the proposed channel model has the ability to reproduce the main properties of real non-stationary channels. Furthermore, the proposed channel model can be adapted to various communication scenarios by adjusting different parameter values.
\end{abstract}


%
\IEEEpeerreviewmaketitle

\section{Introduction}
%
%
%
%
The research and development for the fifth generation (5G) wireless communication networks call for the need for advanced channel models. Considering the fact that the aim of the 5G system is to provide anywhere and anytime connectivity for anyone and anything \cite{5GWangMost,METIS_white_paper,5GNOW,Tullberg20165G,Elayoubi20165G,Shafi20175G}, it requires more general channel models that can simulate a wide range of propagation scenarios such as indoor, urban micro-cell (UMi), urban macro-cell (UMa), suburban macro-cell (SMa), high-speed train (HST), etc.

Another aspect worth considering is the non-stationary properties of channels. Many measurement campaigns and theoretical analyses have shown that in high mobility scenarios, the stationary interval, during which the wide-sense stationary  (WSS) assumption is valid, is much shorter than the observation time interval \cite{StationarityInterval,HSTMeasurement,HSTMeasurement2,HSTMeasurement3,HSTMeasurement4,HST,Wang2016HST,Liu2017tunnel,Liu2017survey}. This means that the WSS condition no longer fulfills in those scenarios.  Channel models relying on the WSS condition may neglect the non-stationarity of the channel and cannot accurately capture certain characteristics of the fast changing channel in high mobility scenarios. As a result, the non-stationary aspects of channels in channel modeling must be carefully taken into account. Besides, most channel models \cite{SCM,SCME,WINNERI,WINNERII,IMTA,WINNERPlus,COST2100} assumed that scatterers are fixed and only the receiver is in motion. However, this is in contrast with real-world moving scenarios, in which the moving scatterers like pedestrians and passing vehicles can have a significant impact on the channel behaviour \cite{Chelli_MovScat,Ali_Chelli_V2V,Karedal_V2V}. Therefore, channel models incorporating fixed and moving scatterers should be studied carefully.

In real wireless propagation environments, scatterers would disperse in a three-dimensional (3-D) space. Measurement data reported in \cite{zhang_3D} illustrated that there is  a wealth of angular information in the vertical plane.  In \cite{zhang_capacity}, the authors proved that the 3-D channel model is more accurate than the two-dimensional (2-D) channel model when calculating the channel capacity, i.e., the assumption of 2-D propagation may lead to an inaccurate estimation of the system performance. Therefore, the 3-D propagation  environment in channel modeling should be considered.

Most standardized channel models, including 3GPP/3GPP2 spatial channel model (SCM)\cite{SCM}, SCM-Extension (SCME)\cite{SCME}, WINNER I\cite{WINNERI}, WINNER II\cite{WINNERII}, and IMT-Advanced (IMT-A) \cite{IMTA} are both 2-D and stationary.
The WINNER+ \cite{WINNERPlus} channel model is a 3-D channel model. It was developed from the widely used 2-D WINNER II channel model by completing the parameter tables with elevation components. The same generic channel model framework was used to model all scenarios, but by using different values of parameters. The WINNER+ channel model mentioned the concept of time evolution but did not elaborate further.
The COST 2100 channel model \cite{COST2100} is a 3-D non-stationary channel model and incorporates time evolution by introducing the concept of visibility region (VR). As the mobile station (MS) enters and leaves different VRs, the corresponding clusters fade in and out. However, the VRs are still 2-D and the clusters of the model are fixed. Besides, the parameters of interesting scenarios cannot easily be extracted through measurement.
In \cite{Ali_Chelli_V2V}, a non-stationary vehicle-to-vehicle (V2V) MIMO channel model using a geometric method was proposed.  The channel model took into account the impact of fixed and moving scatterers. However, it is a 2-D channel model and can only be applied to street scenarios.
The authors in \cite{QuaDRiGa} extended the WINNER channel model with new features that allow the model to support time evolution. But, it can only simulate the propagation environments with fixed clusters.
In \cite{NonIMT-A}, a non-stationary IMT-Advanced MIMO channel model for high-mobility wireless communication systems was proposed. It is able to simulate different moving scenarios by considering small-scale time-variant parameters. However, it is a 2-D channel model that neglected the angle parameters in the vertical plane.
The authors in \cite{YuanYi2015} proposed a 3-D non-stationary MIMO V2V channel model. The parameters are time-variant due to the movements of the transmitter, receiver, and scatterers, which results in a  non-stationary channel model.  The channel model has the ability to study the impacts of vehicular traffic density (VTD) and non-stationarity on
channel statistics. However, it is a pure geometry-based stochastic model (GBSM) which is less versatile than the WINNER family channel models. To the best of the authors' knowledge, a 3-D non-stationary MIMO channel model considering moving clusters and having the ability to simulate a wide range of propagation scenarios is still missing. This paper aims to fill this gap.

In this paper, we extend the stationary channel model in \cite{WINNERPlus} to a non-stationary channel model using a birth-death process and time-variant parameters. The cluster movement is taken into account in the proposed model. The cluster time evolution is redesigned to make the simulation more realistic and enhance the simulation efficiency. Besides the spatial cross-correlation function (CCF), temporal autocorrelation function (ACF), and Doppler power spectral density (PSD), statistical properties such as envelope level-crossing rate (LCR),  envelope average fading duration (AFD), and stationary interval of the proposed channel model are also investigated. By setting proper channel parameters, the proposed channel model can be easily applied to various scenarios or degrade into a 2-D stationary or non-stationary channel model. The major \textbf{contributions} of this paper are summarized as follows:
\begin{enumerate}
\item A novel WINNER+ based 3-D non-stationary wideband MIMO channel model is proposed. Both the receiver and clusters can be moving, which makes the channel model more general. The proposed channel model supports long time/distance simulation and smooth time evolution. By using different parameters in WINNER+ channel model, the proposed channel model can simulate a wide range of propagation scenarios.
\item Statistical properties of the proposed channel model such as spatial CCF,   temporal ACF,  and Doppler PSD are derived and verified by simulations.
\item The  LCR and  AFD of the proposed channel model are derived and compared with the corresponding measurement data. The stationary interval of the proposed channel model is investigated and compared with measurement data and various existing channel models.
\end{enumerate}

The remaining of this paper is organized as follows. In Section II, the proposed channel model is elaborated in detail, which includes geometric construction, channel coefficients generation procedure, and cluster time evolution. Statistical properties such as spatial CCF, temporal ACF, Doppler PSD, LCR,  AFD, and stationary interval are studied in Section III. In Section IV, numerical and  simulation results are presented and analyzed. Conclusions are finally drawn in Section V.

\section{A NOVEL 3-D NON-STATIONARY WIDEBAND MIMO CHANNEL MODEL}
\label{nonstationary_model}

\begin{figure}
\centering\includegraphics[width=3.5in]{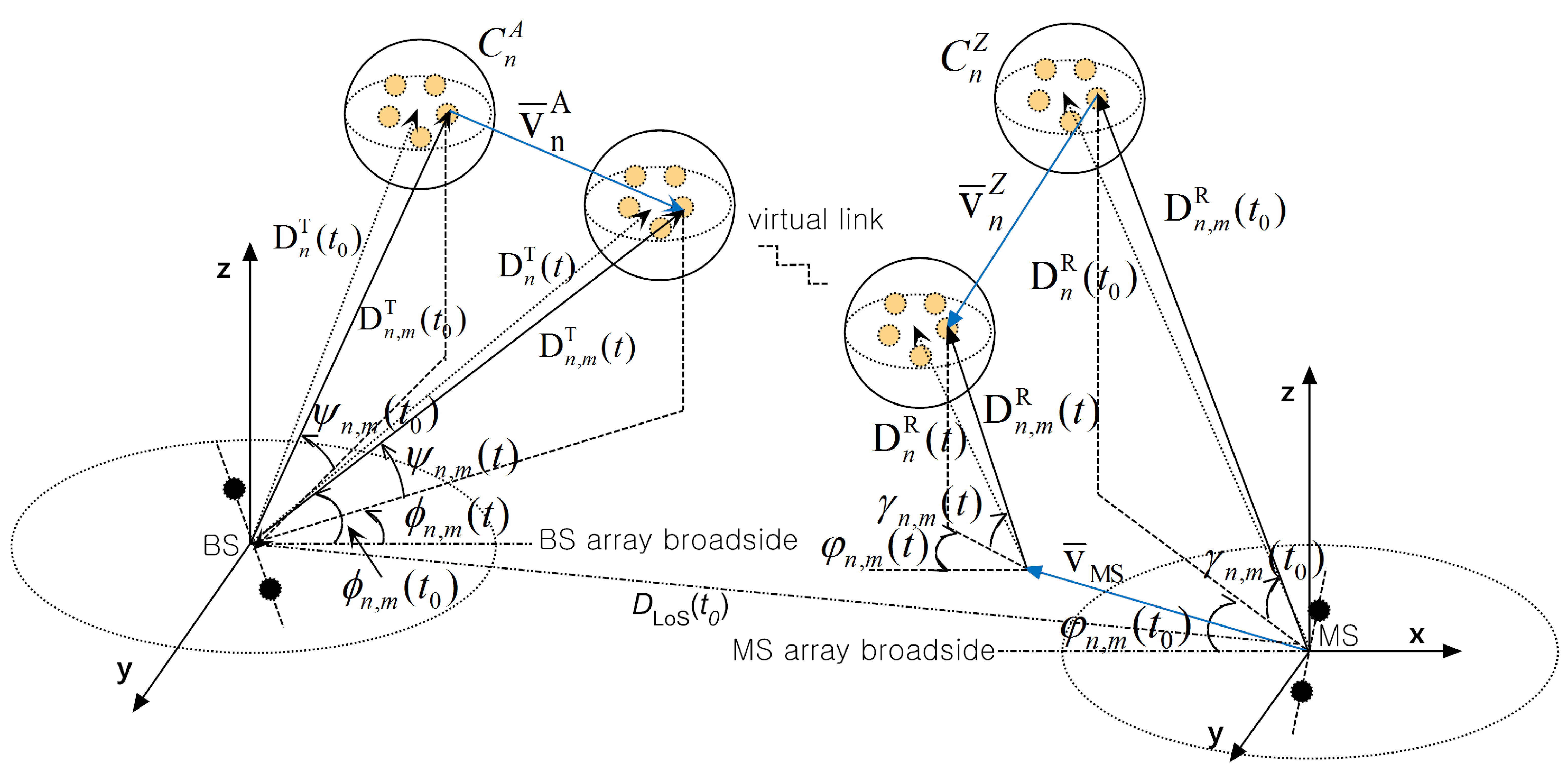}
\caption{BS and MS angle parameters in the proposed 3-D non-stationary channel model.}
\label{fig_angular_parameters}
\end{figure}

The proposed channel model framework is shown in Fig.~\ref{fig_angular_parameters}. For clarity, only the $n$-th ($n=1, ..., N$) path is illustrated. We assume that the base station (BS) is fixed and the MS is in motion. The movement of the MS is defined by a velocity vector $\overline v_\textup{MS}$ with speed $v_\textup{MS}$, travel elevation angle $\vartheta_\textup{MS}$, and travel azimuth angel $\theta_\textup{MS}$. As shown in Fig. \ref{fig_angular_parameters}, each sphere with several dots represents a cluster, which can be in motion or stay static with a scenario dependent probability $P_c$. Considering a multi-bounce scattering propagation, we assume that the angle parameters of rays such as  azimuth angles of departure (AAoDs) and elevation angles of departure (EAoDs) are only related to the first bounce cluster of the $n$-th path, which is denoted by $C^A_n$. Similarly, azimuth angles of arrival (AAoAs) and elevation angles of arrival (EAoAs) are only related to the last bounce cluster of the $n$-th path, which is denoted by $C^Z_n$. The propagation between $C^A_n$ and $C^Z_n$ is not defined. The movement of $C^A_n$ ($C^Z_n$)  is defined by the velocity vector $\overline v^A_n$ ($\overline v^Z_n$) with speed $v^A_n$ ($v^Z_n$), travel elevation angles $\vartheta^A_n$ ($\vartheta^Z_n$), and travel azimuth angles $\theta^A_n$ ($\theta^Z_n$). The moving direction and speed of each cluster are random depending on the specific scenarios. The time-variant parameters including number of clusters, path power, delay, AAoD, AAoA, EAoD, and EAoA which are denoted by $N(t)$, $P_n(t)$, $\tau_n(t)$, $\phi_{n,m}(t)$, $\varphi_{n,m}(t)$, $\psi_{n,m}(t)$, and $\gamma_{n,m(t)}$  are  calculated through the geometrical relationships among transmitter, receiver, and clusters. The subscriber $n,m$ indicates the $m$-th ray of the $n$-th path. The distance between the BS and $C^A_n$, the distance between the MS and $C^Z_n$, and the distance of line-of-sight (LoS) component are denoted by $D_n^T(t)$, $D_n^R(t)$, and $D_{\textup{LoS}}(t)$, respectively. Complete parameters of this model are summarized in Table~\ref{tab_key_geometry}.

\begin{table*}[ht]
\caption{Summary of key parameter definitions.}
\center
    \begin{tabular}{|c|c|}
    \hline
    $C^A_n$ & first bounce cluster of the $n$-th path \\
    \hline
    $C^Z_n$ & last bounce cluster of the $n$-th path \\
    \hline
    $D^T_n(t)$  & distance between the BS and $C^A_n$  \\
    \hline
    $D^R_n(t)$  & distance between the MS and $C^Z_n$ \\
    \hline
    $D^T_{n,m}(t)$  & distance between the BS and $C^A_n$ via the $m$-th ray  \\
    \hline
    $D^R_{n,m}(t)$  & distance between the MS and $C^Z_n$ via the $m$-th ray\\
    \hline
    $D_{\textup{LoS}}(t)$ & distance of the LoS component between the BS and MS  \\
    \hline
    $\phi_{n,m}(t)$, $\varphi_{n,m}(t)$  & azimuth angles of the $m$-th ray of the $n$-th path at the BS and MS sides   \\
    \hline
    $\psi_{n,m}(t)$, $\gamma_{n,m}(t)$  & elevation angles of the $m$-th ray of the $n$-th path at the BS and MS sides   \\
    \hline
    $\phi_{\textup{LoS}}(t)$, $\varphi_{\textup{LoS}}(t)$  & azimuth angles of the LoS component at the BS and MS sides     \\
    \hline
    $\psi_{\textup{LoS}}(t)$, $\gamma_{\textup{LoS}}(t)$  & elevation angles of the LoS component at the BS and MS sides   \\
    \hline
    $\overline v^A_n$, $\overline v^Z_n$, $\overline v_{\textup{MS}} $ & velocity vectors of $C^A_n$, $C^Z_n$, and the MS  \\
    \hline
    $v^A_n, \theta^A_n, \vartheta ^A_n$ & speed, travel azimuth angle, and travel elevation angle of $C^A_n$\\
    \hline
    $v^Z_n, \theta^Z_n, \vartheta ^Z_n$ & speed, travel azimuth angle, and travel elevation angle of $C^Z_n$\\
    \hline
    $v_\textup{MS}, \theta_{\textup{MS}}, \vartheta _{\textup{MS}}$ & speed, travel azimuth angle, and travel elevation angle of the MS\\
    \hline
    \end{tabular}
    \label{tab_key_geometry}
\end{table*}

\begin{figure}
\centering\includegraphics[ height = 2.8in, width=3.3in]{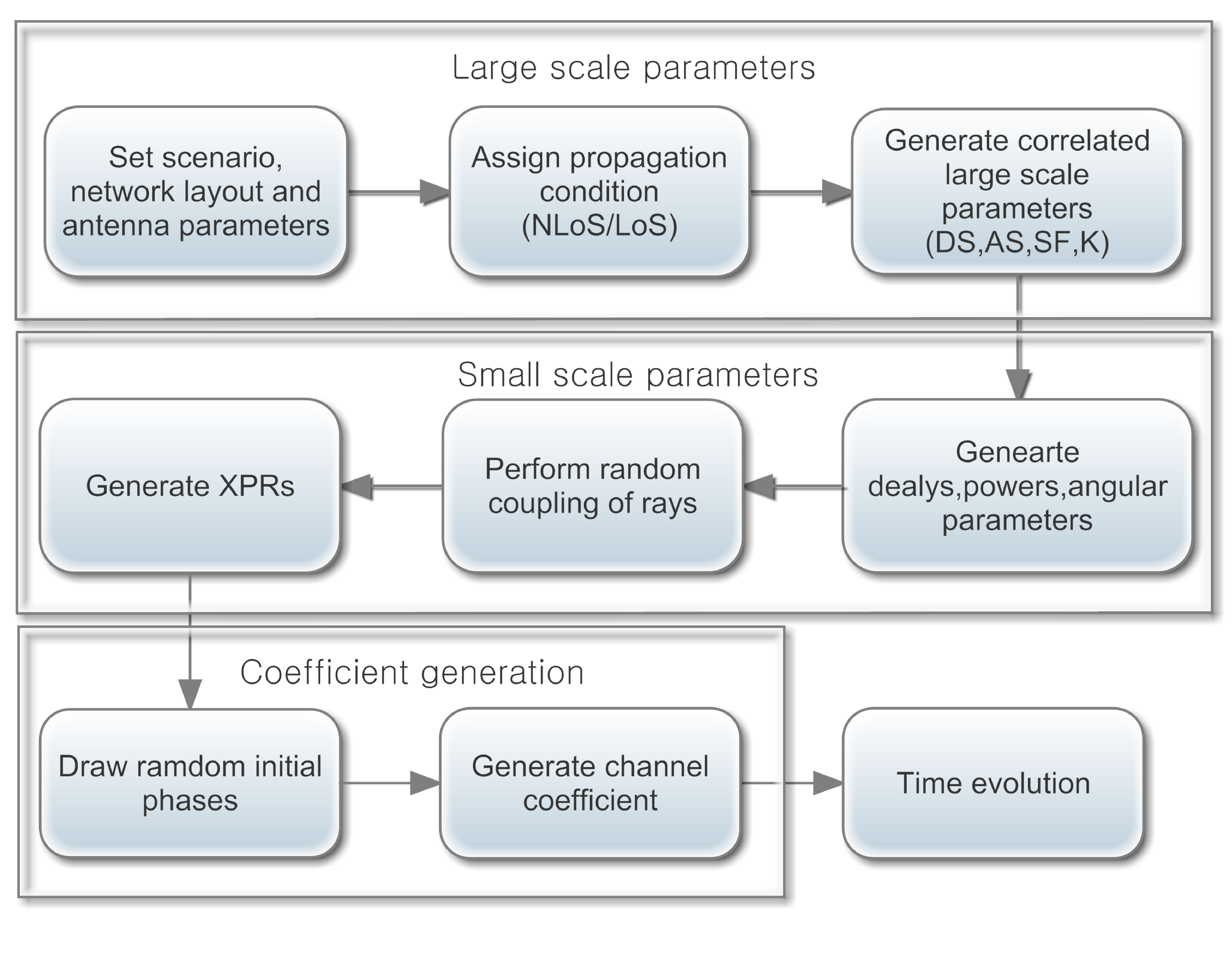}
\caption{Channel coefficients initialization.}
\label{fig_LSP_SSP}
\end{figure}

The generation process of the proposed channel model can be divided into two stages, i.e.,  initialization and  time evolution, which are shown in Fig. \ref{fig_LSP_SSP}. The first stage includes the large scale parameters (LSPs) generation and  small scale parameters (SSPs) generation \cite{WINNERPlus}. We will focus on the time evolution stage, which will be introduced in the rest of this paper.

\subsection{Channel Impulse Response (CIR) }\label{CIR}
The channel coefficients from transmit antenna element $s$ to receive antenna element $u$ via the $n$-th cluster (consisting $M$ rays) are the superposition of the LoS component and non-line-of-sight (NLoS) components, and can be expressed as
\begin{equation}
h_{u,s,n}(t)=h^\text{LoS}_{u,s,n}(t)+h^\text{NLoS}_{u,s,n}(t)
\label{equ_CIR}
\end{equation}
where
\begin{align}
    &h_{u,s,n}^\text{NLoS}(t)=\sqrt{\frac{P_{n}(t)}{(K+1)M}}\sum_{m=1}^{M}\begin{bmatrix}
    F_{rx,u,V}(\gamma_{n,m},\varphi_{n,m})\\
    F_{rx,u,H}(\gamma_{n,m},\varphi_{n,m})
    \end{bmatrix}^\text{T} \nonumber \\
    &\begin{bmatrix}
    \sqrt{\kappa_{n,m}^{-1}}e^{j\Theta_{n,m}^{VV}} & e^{j\Theta_{n,m}^{VH}}\\
    e^{j\Theta_{n,m}^{HV}} & \sqrt{\kappa_{n,m}^{-1}}e^{j\Theta_{n,m}^{HH}}
    \end{bmatrix}
    \begin{bmatrix}
    F_{tx,s,V}(\psi_{n,m},\phi_{n,m})\\
    F_{tx,s,H}(\psi_{n,m},\phi_{n,m})
    \end{bmatrix}\nonumber\\
    &\cdot e^{j2\pi\lambda ^{-1}\overline{r}_s\cdot \overline{\Phi }_{n,m}(t)}\cdot e^{j2\pi\lambda ^{-1}\overline{r}_u\cdot \overline{\Psi  }_{n,m}(t)}\cdot e^{j2\pi\nu _{n,m}(t)\cdot t}
\label{CIR_NLoS}
\end{align}
and
\begin{align}
h^\text{LoS}_{u,s,n}(t)=&
\delta (n-1)\cdot \sqrt{\frac{K}{K+1}}\cdot \begin{bmatrix}
F_{rx,u,V}(\gamma_{\textup{LoS}},\varphi_{\textup{LoS}})\\
F_{rx,u,H}(\gamma_{\textup{LoS}},\varphi_{\textup{LoS}})
\end{bmatrix}^\textup{T} \nonumber \\
&\begin{bmatrix}
e^{j\Theta_{\textup{LoS}}^{VV}} & 0\\
0 & e^{j\Theta_{\textup{LoS}}^{HH}}
\end{bmatrix}
\begin{bmatrix}
F_{tx,s,V}(\psi_{\textup{LoS}},\phi_{\textup{LoS}})\\
F_{tx,s,H}(\psi_{\textup{LoS}},\phi_{\textup{LoS}})
\end{bmatrix}\nonumber\\
&\cdot e^{j2\pi\lambda ^{-1}\overline{r}_s\cdot \overline{\Phi }_{\textup{LoS}}(t)}\cdot e^{j2\pi\lambda ^{-1}\overline{r}_u\cdot \overline{\Psi  }_{\textup{LoS}}(t)}\cdot e^{j2\pi\nu _{\textup{LoS}}(t)\cdot t}
\label{CIR_LoS}.
\end{align}

For the NLoS components (2),  $K$ is the Rice factor, $P_n(t)$ is the time-variant power of the $n$-th cluster, $F_{rx,u,V}$ and $F_{rx,u,H}$ are the radiation patterns of antenna element $u$ for vertical and horizontal polarizations, respectively, $F_{tx,s,V}$ and $F_{tx,s,H}$ are the radiation patterns of antenna element $s$ for vertical and horizontal polarizations, respectively. Here, $\Theta_{n,m}^{VV}$, $\Theta_{n,m}^{VH}$, $\Theta_{n,m}^{HV}$, and $\Theta_{n,m}^{HH}$ are uniformly distributed random initial phases of the $m$-th ray of the $n$-th cluster for four different polarization combinations, $\kappa_{n,m}$ is the cross polarization power ratio, and $\overline{\Phi}_{n,m}(t)$ is the time-variant departure angle unit vector of the $m$-th ray of the $n$-th cluster with EAoD $\psi_{n,m}(t)$  and AAoD $\phi_{n,m}(t)$, and can be expressed as
\begin{equation}
\overline{\Phi}_{n,m}(t) =\begin{bmatrix}
 \textup{cos}\psi _{n,m}(t)\cdot \textup{cos}\phi_{n,m}(t)
\\ \textup{cos}\psi _{n,m}(t)\cdot \textup{sin}\phi_{n,m}(t)
\\ \textup{sin}\psi_{n,m}(t)
\end{bmatrix}^\textup T.
\end{equation}
In (\ref{CIR_NLoS}), $\overline{\Psi}_{n,m}(t)$ is the time-variant arrival angle unit vector of the $m$-th ray of the $n$-th cluster with EAoA $\gamma _{n,m}(t)$ and AAoA $\varphi _{n,m}(t)$, and can be given by
\begin{equation}
\overline{\Psi}_{n,m}(t)=\begin{bmatrix}\textup{cos}\gamma _{n,m}(t)\cdot \textup{cos}\varphi_{n,m}(t)
\\ \textup{cos}\gamma _{n,m}(t)\cdot \textup{sin}\varphi_{n,m}(t)
\\ \textup{sin}\gamma_{n,m}(t)
\end{bmatrix}^\textup T.
\end{equation}
The coordinates of antenna element $s$ and antenna element $u$, i.e. $\overline{r}_s=(x_s, y_s, z_s)$ and $\overline{r}_u=(x_u, y_u, z_u)$, are local coordinate systems (LCSs) relative to the first antenna elements of the transmit array and receive array, respectively. The MS movement denoted by the MS velocity vector $\overline v_{\textup{MS}}$ can be expressed as
\begin{equation}
\overline v_{\textup{MS}}=v_{\textup{MS}}\cdot \begin{bmatrix}
\textup{cos}\vartheta_{\textup{MS}}\cdot \textup{cos}\theta_{\textup{MS}}\\ \textup{cos}\vartheta_{\textup{MS}}\cdot \textup{sin}\theta_{\textup{MS}}\\ \textup{sin}\vartheta_{\textup{MS}}
\end{bmatrix}^\textup T.
\end{equation}
Furthermore, the time-variant Doppler frequency  $\nu_{n,m}(t)$ containing the effects of the moving MS and moving scatterers can be expressed as
\begin{align}
\nu _{n,m}(t)=(\overline v_{\textup{MS}}\cdot\overline\Psi_{n,m}(t)-\overline v_n^A\cdot\overline\Phi_{n,m}(t)-\overline v_n^Z\cdot\overline\Psi_{n,m}(t))/\lambda
\label{equ_dop}
\end{align}
where $\lambda$ is the carrier wavelength, $\overline v_{\textup{MS}}\cdot\overline\Psi_{n,m}(t)/\lambda$ is the Doppler frequency caused by the moving MS, $-\overline v_n^A\cdot\overline\Phi_{n,m}(t)/\lambda$ and $-\overline v_n^Z\cdot\overline\Psi_{n,m}(t)/\lambda$ are the Doppler frequencies generated due to transmitted wave impinging on the moving scatterer $C_n^A$ and the moving scatterer $C_n^Z$ redirecting the wave to the receiver, respectively.

For the LoS component (\ref{CIR_LoS}), $\delta(\cdot)$ denotes the Dirac function,
$\overline{\Phi }_{\textup{LoS}}(t)$ is the time-variant departure angle unit vector of the LoS path with EAoD $\psi_\text{LoS}(t)$ and AAoD $\phi_\text{LoS}(t)$, $\overline{\Psi }_{\textup{LoS}}(t)$ is the time-variant arrival angle unit vector of the LoS path with AAoA $\gamma_\text{LoS}(t)$ and EAoA $\varphi_\text{LoS}(t)$, $\Theta_{\textup{LoS}}^{VV}$ and $\Theta_{\textup{LoS}}^{HH}$ are the random initial phases of the LoS path for $\{vv,hh\}$ polarization combinations, respectively, and $\nu _{\textup{LoS}}(t)$ is the time-variant Doppler frequency of the LoS component caused by the MS and scatterer movement, and can be given by
\begin{equation}
\nu _\text{LoS}(t)=(\overline v_{\textup{MS}}\cdot\overline\Psi_\text{LoS}(t)-\overline v_n^A\cdot\overline\Phi_\text{LoS}(t)-\overline v_n^Z\cdot\overline\Psi_\text{LoS}(t))/\lambda.
\end{equation}

%

The CIR of the proposed channel model between transmit antenna element $s$ and receive antenna element $u$ can be expressed as
\begin{equation}
h_{u,s}(t,\tau)=\sum_{n=1}^{N(t)}h_{u,s,n}(t)\cdot\delta(\tau-\tau_n(t)).
\end{equation}

\subsection{Time Evolution }\label{time_evolution}
 \begin{figure}
\centering\includegraphics[height=4in,width=2.5in]{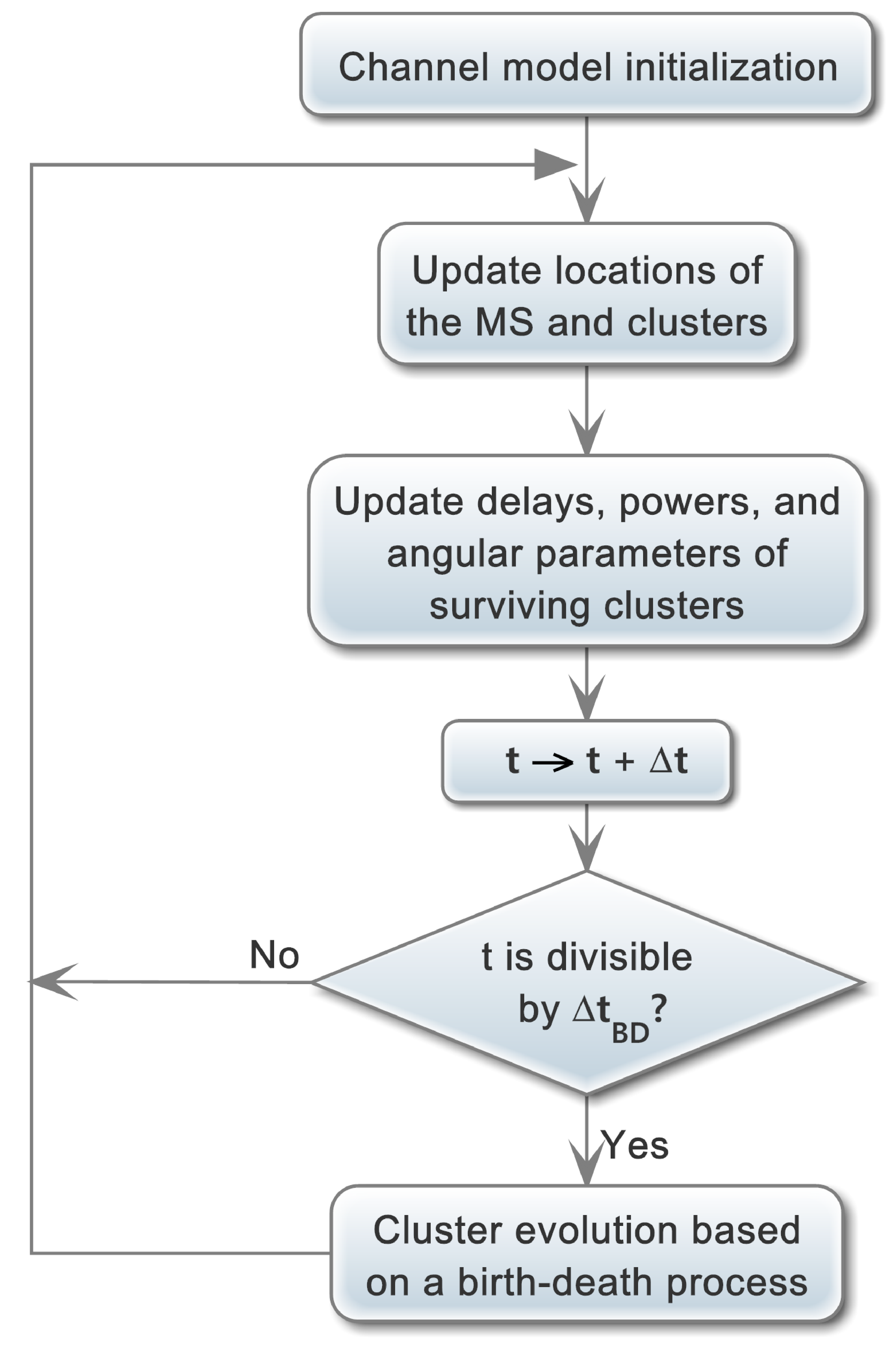}
\caption{Flowchart of cluster evolution.}
\label{flowchart}
\end{figure}

The non-stationarity of the proposed channel model is embodied in two mechanisms, i.e., the time-variant parameters and the birth-death process \cite{NonIMT-A,BirthDeath}. The time-variant parameters should be updated constantly, while the clusters in a specific scenario can exist over a certain time period, which means the number of the clusters do not change frequently. To combine these two mechanisms to achieve a smooth time evolution and enhance simulation efficiency, we introduce two different sampling intervals, i.e., channel sampling interval $\Delta t$ and birth-death sampling interval $\Delta t_\textup{BD}$. During $\Delta t$, the parameters such as angles, powers, and delays are updated. While, during the $\Delta t_\textup{BD}$, the cluster birth and death occurs. Note that $\Delta t_\textup{BD}$ is an integer multiple of $\Delta t$. The time evolution flowchart is shown in Fig. \ref{flowchart}.

The time variance of a wireless channel can be caused by not only the movement of the MS, but also the moving scatterers, e.g., passing vehicles and pedestrians. The variable $\delta_{P}(t,\Delta t_\textup{BD})$ is introduced here to describe how much the propagation environment varies during the time interval between $t$ and $t+\Delta t_\textup{BD}$ and can be used as a measure of the channel fluctuation. Therefore, the channel fluctuation during the time interval between $t$ and $t+\Delta t_\textup{BD}$ can be defined as
 \begin{equation}\label{equ_ChannelFluctuation}
 \delta_{P}(t,\Delta t_\textup{BD})=\delta_{\textup{MC},n}(t,\Delta t_\textup{BD})+\delta_{\textup{MS}}(t,\Delta t_\textup{BD})
 \end{equation}
where $\delta_{\textup{MC},n}(t,\Delta t_\textup{BD})$ is the channel fluctuation resulting from the movement of scatterers and can be defined as
 \begin{equation}
 \delta_{\textup{MC},n}(t,\Delta t_\textup{BD})=\int_{t}^{t+\Delta t_\textup{BD}}P_c\cdot(\left | \overline v^A_n(t)  \right |+\left | \overline v^Z_n(t)  \right |)dt
 \label{equ_MC_move}
 \end{equation}
where $P_c$ is the probability of cluster movements. In (\ref{equ_ChannelFluctuation}), the variable $ \delta_{\textup{MS}}(t,\Delta t_\textup{BD})$ is the channel fluctuation resulting from the movement of the MS and can be defined as
 \begin{equation}
 \delta_{\textup{MS}}(t,\Delta t_\textup{BD})=\int_{t}^{t+\Delta t_\textup{BD}}(\left | \overline v_{\textup{MS}}(t)  \right |)dt
 \label{equ_MS_move}.
 \end{equation}
Because of short time intervals, we assume that each cluster and the MS travel at constant speeds in the time interval $\Delta t_\textup{BD}$ and mean velocities of clusters are used in this process, i.e., $v^A=\textup{E}[v_n^A]$, $v^Z=\textup{E}[v_n^Z]$. Therefore, (\ref{equ_MC_move}) and (\ref{equ_MS_move}) can be simplified to
\begin{equation}
\delta_{\textup{MC}}(\Delta t_\textup{BD})=P_c\cdot(v^A+v^Z)\cdot \Delta t_\textup{BD}
\end{equation}
and
\begin{equation}
\delta_{\textup{MS}}(\Delta t_\textup{BD})=v_{\textup{MS}}\cdot \Delta t_\textup{BD}.
\end{equation}
Note that, all the clusters share the same survival probability and $\delta _P(\Delta t_\textup{BD})$ can be used as a driving parameter for the recombination and generation processes of multi-path components.

According to the birth-death process, the clusters at time instant $t+\Delta t_\textup{BD}$ can be assumed as the sum of surviving clusters that already exist at time instant $t$ and the newly generated clusters during the time interval  $\Delta t_\textup{BD}$. The process is determined by a generation rate of clusters $\lambda_G$ and a  recombination rate of clusters $\lambda_R$. The expectation of the total number of clusters in the proposed channel model can be calculated as
\begin{equation}
\textup{E}[N(t)]=\frac{\lambda_G}{\lambda_R}.
\end{equation}
The probabilities of clusters at $t+\Delta t_\textup{BD}$ survived from $t$ can be modeled as
\begin{equation}
P_{\textup{survival}}(\Delta t_\textup{BD})=e^{-\lambda_R\cdot\frac{\delta _P(\Delta t_\textup{BD})}{D_c}}
\end{equation}
where $D_c$ is the scenario dependent correlation factor. According to Poisson process, the durations between clusters appearance and disappearance follow exponential distribution. The expectation of the number of newly generated clusters can be computed as
\begin{equation}
\textup{E}[N_{\textup{new}}(t+\Delta t_\textup{BD})]=\frac{\lambda_G}{\lambda_R}(1-e^{-\lambda_R \cdot\frac{\delta _P(\Delta t_\textup{BD})}{D_c}}).
\end{equation}

As is shown in Fig. \ref{flowchart}, the disappearing clusters at each time instant are removed from the channel model. For the newly generated clusters, delays, path powers, and angle parameters are randomly generated which are similar to the initialization process at time $t_0$. For the surviving clusters from the previous time instant, the update process of  delays, powers and angle parameters will be described in  rest of this section. Note that the correlation between a cluster and its successor is quantified by the movement variable $\delta_{P}(\Delta t_\text{BD})$, higher values of $\delta_{P}(\Delta t_\text{BD})$ will lead to a weak correlation between the properties of  ancestor clusters at time instant $t$ and their successor at time instant $t+\Delta t_\text{BD}$, which means the channel changes rapidly. On the contrary, lower values of $\delta_{P}(\Delta t_\text{BD})$ indicate the properties of clusters between time instant $t$ and $t+\Delta t_\text{BD}$ are strongly correlated, which results in a slow changing channel.

In order to achieve a smooth transition when a cluster appear/disappear, the transition region \cite{Correia2006Mobile,Verdone2012Pervasive} is adopted to make a cluster fade in and fade out smoothly. We extend \cite[eq.~(3.35)]{Verdone2012Pervasive} by incorporating the whole lifetime of a cluster, i.e., fading in, existing over a certain time period, and fading out. The power of the $n$-th cluster is controlled by the attenuation factor $\xi_n(t)$, which can be expressed as
\begin{equation}
\xi_n(t)=\frac{1}{2}-\frac{1}{\pi}\cdot\textup{arctan}(\frac{2[L_c+(\left | 2t-T_n\right | - T_n)\cdot v_\textup{MS})]}{\sqrt{\lambda\cdot L_c}}), \nonumber
\end{equation}
\begin{equation}
t\in \left [ 0,T_n \right ]  \hspace{-5.5cm}
\end{equation}
where $T_n$ is the lifetime of the $n$-th cluster, $L_c$ is the length of the transition region.
\begin{figure}
\centering\includegraphics[width=3.5in]{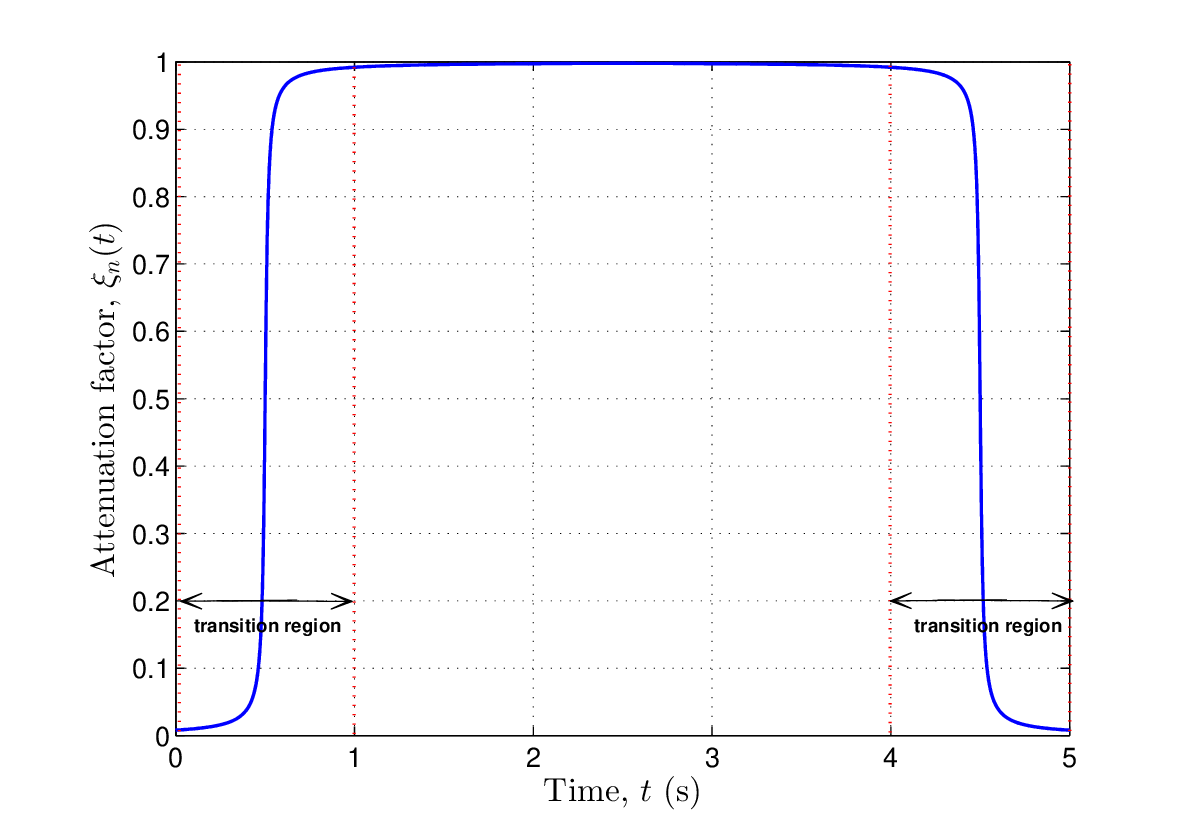}
\caption{Attenuation factor ($v_\textup{MS}$ = 60 m/s, $T_n$ = 5 s, $L_c$ = 60 m).}
\label{fig_transition_region}
\end{figure}
\begin{figure}
\centering\includegraphics[width=3.5in]{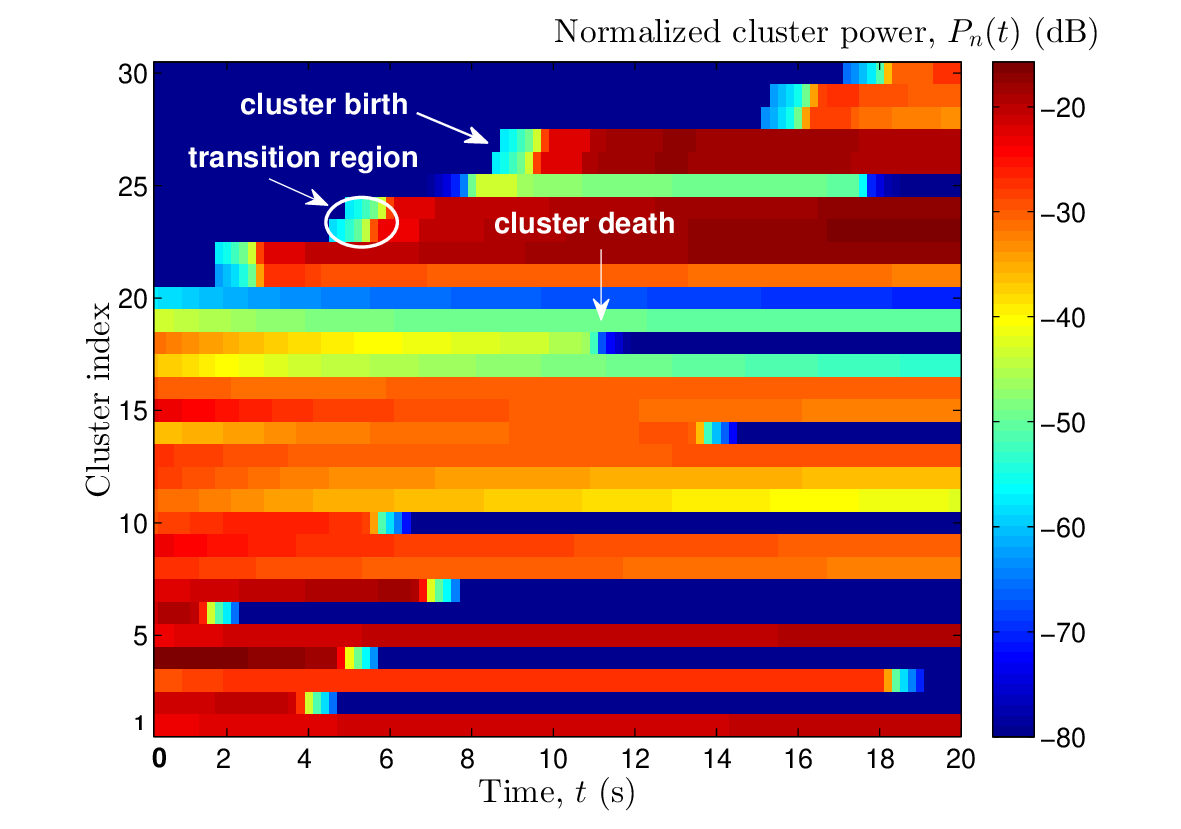}
\caption{Time-variant clusters with birth death process ($N(t_0)$ = 20, $\lambda_G$ = 0.8 m, $\lambda_R$ = 0.04 m, $v_\textup{MS}$ = 60 m/s, $ v^A = 15$ m/s, $v^Z = 5$ m/s, $P_c$ = 0.3, $D_c$ = 10 m, $L_c$ = 60 m).}
\label{fig_BirthDeath}
\end{figure}

The attenuation factor is depicted in Fig. \ref{fig_transition_region}. The cluster lifetime is assumed to be 5 s. A smooth behaviour can be observed in the transition region. Fig. \ref{fig_BirthDeath} illustrates the clusters evolution in UMa scenario with NLoS case. The initial number of clusters $N(t_0)=20$. It shows the newly born clusters fade in and dying clusters fade out according to the birth-death process.
\subsection{Evolution of Surviving Clusters}\label{evolution_of_surviving_clusters}
\subsubsection{Update moving clusters (MCs) and MS locations }\label{loacations}

Since the BS is fixed, the distance between BS and $C^A_n$ only depends on the movements of $C^A_n$. Let  $D^T_{n,m}(t)$ denotes  the distance  between BS and  $C^A_n$ via the $m$-th ray at time $t$. We assume that the initial diatance $D^T_{n,m}(t_0)$  follows a Gaussian distribution, i.e., $D^T_{n,m}(t_0)\sim N(D^T_{n}(t_0),\sigma _D)$, where $\sigma _D$ is distance spread of the clusters depends on specific scenarios. As is shown in Fig. \ref{fig_angular_parameters},
let $\overline D^T_{n,m}(t_0)$ and $\overline D^T_{n,m}(t)$ denote the position vectors from BS to the $C^A_n$ via the $m$-th ray at time $t_0$ and time $t$, respectively.  $\overline D^T_{n,m}(t)$  can be calculated by
\begin{equation}
\overline D^ T_{n,m}(t)=\overline D^ T_{n,m}(t_0)+ \overline v^A_n \cdot t
\label{vec_D_T}
\end{equation}
where
\begin{equation}
\overline D^ T_{n,m}(t_0)=D^ T_{n,m}(t_0) \cdot \overline \Phi(t_0)
\end{equation}
 and
\begin{equation}
\overline v^A_n=v^A_n \cdot \begin{bmatrix}\textup{cos}\vartheta^A_n\cdot \textup{cos}\theta^A_n,\\ \textup{cos}\vartheta^A_n\cdot \textup{sin}\theta^A_n,\\ \textup{sin}\vartheta^A_n\end{bmatrix}^\textup T.
\end{equation}
The distance between BS and $C^A_n$ at time $t$ can be calculated by
\begin{equation}
D^ T_{n}(t)=\frac{1}{M}\sum_{m=1}^{M}\left| \overline D^ T_{n,m}(t) \right|
\label{equa_D_T}.
\end{equation}

The distance between MS and $C^Z_n$ depends on the relative motion between  MS and $C^Z_n$. $\overline D^R_{n,m}(t_0)$ and $\overline D^R_{n,m}(t)$ are the position vectors from MS to the $C^Z_n$ at time $t_0$ and time $t$, respectively. Similarly, $D^R_{n,m}(t_0)\sim N(D^R_n(t_0),\sigma _D)$.  The $\overline D^R_{n,m}(t)$ is calculated as
\begin{align}
\overline D^R_{n,m}(t)=  \overline D^R_{n,m}(t_0)+\overline{v}^Z_n \cdot t -\overline {v}_{\textup{MS}}\cdot t
\label{vec_D_R}
\end{align}
where
\begin{equation}
\overline D^R_{n,m}(t_0)=D^R_{n,m}(t_0)\cdot  \overline\Psi(t_0)
\end{equation}
and
\begin{equation}
\overline v^Z_n=v^Z_n \cdot \begin{bmatrix}\textup{cos}\vartheta^Z_n\cdot \textup{cos}\theta^Z_n \\ \textup{cos}\vartheta^Z_n\cdot sin\theta_Z  \\ \textup{sin}\vartheta^Z_n\end{bmatrix}^\textup T.
\end{equation}
The distance between MS and the $C^Z_n$ at time $t$ can be calculated as
\begin{equation}
D^R_n(t)=\frac{1}{M}\sum_{m=1}^{M}\left| \overline D^R_{n,m}(t) \right|
\label{equa_D_R}.
\end{equation}
In addition, the distance between BS and MS at time $t$ can be calculated as
\begin{equation}
D_\textup{LoS}(t)=\left| \overline D_\textup{LoS}(t_0) +\overline{v}_{\textup{MS}}\cdot t\right|
\label{equa_D_LoS}
\end{equation}
where
\begin{equation}
\overline D_\textup{LoS}(t_0) = D_\textup{LoS}(t_0) \cdot \begin{bmatrix}\textup{cos}\psi_\textup{LoS}(t_0)\cdot \textup{cos} \phi_\textup{LoS}(t_0)\\
\textup{cos}\psi_\textup{LoS}(t_0)\cdot \textup{sin} \phi_\textup{LoS}(t_0) \\
\textup{sin}\psi_\textup{LoS}(t_0)
\end{bmatrix}^\textup T.
\end{equation}

\subsubsection{Update delays, $\tau _n(t)$}
The time-variant $D^T_n(t)$, $D^R_n(t)$, and $D_\textup{LoS}(t)$ have been obtained as (\ref{equa_D_T}), (\ref{equa_D_R}), and (\ref{equa_D_LoS}) using a geometric method. We assume that the delay of the $n$-th path at time $t$ which is denoted by $\tau_n(t)$ consists of three parts: the delay of  the first bounce, the delay of the last bounce, and the delay of the virtual link  between $C^A_n$ and $C^Z_n$. Therefore, $\tau_n(t)$ can be calculated as
\begin{equation}
\tau_n(t)=\frac{D^T_n(t)+D^R_n(t)}{c}+\tilde{\tau}_n(t)
\end{equation}
where $c$ is the speed of light. The delay of the virtual link $\tilde{\tau}_n(t)$ can be calculated  according to a first-order filtering algorithm \cite{BirthDeath3}:
\begin{equation}
\tilde{\tau}_n(t)=e^{-{\frac{\Delta t}{\zeta}}}\cdot \tilde \tau_n(t-\Delta t)+(1-e^{-{\frac{\Delta t}{\zeta}}})\cdot X
\end{equation}
where $X$ is drawn randomly with uniform distribution, i.e., $X\sim U(D_\textup{LoS}(t)/c,\tau_\textup{max})$ and $\tau_\textup{max}$ can be found in \cite{IMTA}. $\zeta$ is a parameter which depends on the coherence of a virtual link and specific scenarios. The delay of virtual link at initial time $t_0$ can be calculated as
\begin{equation}
\tilde\tau_n(t_0)=\tau _n(t_0)-\frac{D^T_n(t_0)+D^R_n(t_0)}{c}.
\end{equation}
The initial delays $\tau_n(t_0)$  are randomly generated with exponential delay distribution or uniform distribution in UMi scenario with NLoS condition\cite{WINNERII}.

\subsubsection{Update powers of clusters, $P_n(t)$}
The cluster power can be calculated through the time-variant delay $\tau_n(t)$ with a single slope exponential power delay profile and the attenuation factor $\xi(t)$. The cluster power is determined by
\begin{equation}
P'_n(t)=\left | \xi_n(t) \right |^2 \cdot e^{-\tau_n(t)\cdot\frac{r_\tau-1}{r_\tau\sigma _\tau}}\cdot10^{\frac{-Z_n}{10}}
\end{equation}
and for the NLoS UMi scenario, the cluster powers is determined by
\begin{equation}
P'_n(t)=\left | \xi_n(t) \right |^2 \cdot e^{\frac{-\tau_n(t)}{\sigma_\tau}}\cdot10^{\frac{-Z_n}{10}}
\end{equation}
where $r_\tau$ is the delay distribution proportionality factor, $\sigma _\tau$ is the delay spread, $Z_n$ is the per cluster shadowing term in dB. The powers are normalized so that the total power of all clusters is equal to one. Note that the total power of clusters at time instant $t$ contains the power of survived clusters from the previous time instant and the power of newly generated clusters. The normalization process can be expressed as
\begin{equation}
P_n(t)=\frac{P'_n(t)}{\sum_{n=1}^{N(t)}P'_n(t)}.
\end{equation}

%

\subsubsection{Update angle parameters}
The position vector $\overline D^T_{n,m}(t)$  from BS to  $C^A_n$ has been obtained through ({\ref{vec_D_T}). The x, y, and z  components of $\overline D^T_{n,m}(t)$, i.e., $x^T_{n,m}(t)$, $y^T_{n,m}(t)$, and $z^T_{n,m}(t)$ can be calculated as

\begin{align}
x^T_{n,m}(t) =& D^T_{n,m}(t_0)\cdot \textup{cos}\psi_{n,m}(t_0)\cdot \textup{cos}\phi _{n,m}(t_0)+\nonumber \\
 &v^A_n \cdot \textup{cos}\vartheta^A_n\cdot \textup{cos}\theta^A_n\cdot t \\
y^T_{n,m}(t) =& D^T_{n,m}(t_0)\cdot \textup{cos}\psi_{n,m}(t_0)\cdot \textup{sin}\phi_{n,m}(t_0)+\nonumber \\
 &v^A_n\cdot \textup{cos}\vartheta^A_n\cdot \textup{sin}\theta^A_n \cdot t \\
z^T_{n,m}(t) =&D^T_{n,m}(t_0)\cdot \textup{sin}\psi_{n,m}(t_0)+v^A_n\cdot \textup{sin}\vartheta^A_n\cdot t.
\end{align}
The position vector $\overline D^R_{n,m}(t)$ from MS to $C^Z_n$ has been calculated through  ({\ref{vec_D_R}). Similarly, the three axes components of $\overline D^R_{n,m}(t)$, i.e.,  $x^R_{n,m}(t)$, $y^R_{n,m}(t)$, and $z^R_{n,m}(t)$ can be calculated as

\begin{align}
x^R_{n,m}(t)=&D^R_{n,m}(t_0)\cdot \textup{cos}\gamma_{n,m}(t_0)\cdot \textup{cos}\varphi _{n,m}(t_0)+\nonumber \\&v^Z_n\cdot \textup{cos}\vartheta^Z_n\cdot \textup{cos}\theta^Z_n\cdot t-v_\textup{MS}\cdot \textup{cos}\vartheta_\textup{MS}\cdot \textup{cos}\theta_\textup{MS} \cdot t \\
y^R_{n,m}(t)=&D^R_{n,m}(t_0)\cdot \textup{cos}\gamma_{n,m}(t_0)\cdot \textup{sin}\varphi_{n,m}(t_0)+\nonumber \\&v^Z_n\cdot \textup{cos}\vartheta^Z_n\cdot \textup{sin}\theta^Z_n\cdot t-v_\textup{MS}\cdot \textup{cos}\vartheta_\textup{MS}\cdot \textup{sin}\theta_\textup{MS}\cdot t \\
z^R_{n,m}(t)=&D^R_{n,m}(t_0)\cdot \textup{sin}\gamma_{n,m}(t_0)+v^Z_n\cdot \textup{sin}\vartheta^Z_n\cdot t-\nonumber \\ &v_\textup{MS}\cdot \textup{sin}\vartheta_\textup{MS}\cdot t.
\end{align}
 Therefore, the time-variant EAoDs and EAoAs  can be calculated by transform the Cartesian coordinates into spherical coordinates:
 \begin{equation}
 \psi_{n,m}(t)=\textup{arcsin}(\frac{z^T_{n,m}(t)}{D^T_{n,m,}(t)})
 \end{equation}
 and
 \begin{equation}
 \gamma_{n,m}(t)=\textup{arcsin}(\frac{z^R_{n,m}(t)}{D^R_{n,m,}(t)}).
 \end{equation}
 Similarly, the time-variant AAoDs and AAoAs can be calculated as

\begin{align}
\phi_{n,m}(t)\!=\!\left\{\begin{aligned}
&\phi'_{n,m}(t),\ \ x^T_{n,m}(t)\geqslant 0\\
&\phi'_{n,m}(t)+180^{\circ},\ x^T_{n,m}(t)<0\ \textup{and}\  y^T_{n,m}(t)\geqslant 0 \\
&\phi'_{n,m}(t)-180^{\circ},\ x^T_{n,m}(t)<0\ \textup{and}\  y^T_{n,m}(t)<0 \\
\end{aligned}\right.
\end{align}
where
 \begin{equation}
 \phi'_{n,m}(t) = \textup{arctan}(\frac{y^T_{n,m}(t)}{x^T_{n,m}(t)})
 \end{equation}
 and

\begin{align}
\varphi_{n,m}(t)\!=\!\left\{\begin{aligned}
&\varphi'_{n,m}(t),\ \ x^R_{n,m}(t)\geqslant 0\\
&\varphi'_{n,m}(t)+180^{\circ},\ x^R_{n,m}(t)<0\ \textup{and}\  y^R_{n,m}(t)\geqslant 0 \\
&\varphi'_{n,m}(t)-180^{\circ},\ x^R_{n,m}(t)<0\ \textup{and}\  y^R_{n,m}(t)<0 \\
\end{aligned}\right.
\end{align}
where
 \begin{equation}
 \varphi'_{n,m}(t) = \textup{arctan}(\frac{y^R_{n,m}(t)}{x^R_{n,m}(t)}).
 \end{equation}

The angle parameters of the LoS path are updated using the similar process.

\section{STATISTICAL PROPERTIES OF 3-D NON-STATIONARY WIDEBAND MIMO CHANNEL MODEL} \label{statistical_properties}

\subsection{Spatial CCF}
The normalized complex spatial-temporal correlation function between two channel coefficients $h_{u_1,s_1,n}$ and $h_{u_2,s_2,n}$ is defined as \cite{ACFCCF2}
\begin{align}
\rho_{(s_1, u_1), (s_2, u_2), n}(t,\Delta t, \Delta d_s,\Delta d_u)\nonumber \\=\textup E
\begin{Bmatrix}
\frac{h_{u_1,s_1,n}(t)\cdot{h^*_
{u_2,s_2,n}}(t+\Delta t)}{\sigma_{h_{u_1,s_1,n}}\cdot \sigma_{h_{u_2,s_2,n}} }
\end{Bmatrix}
\label{equ_spa_tem}
\end{align}
where $\textup E\{\cdot\}$ denotes the statistical average and $(\cdot)^*$ denotes the complex conjugation operation. $\Delta d_s= \left| d_{s_1}-d_{s_2} \right|$ and  $\Delta d_u= \left| d_{u_1}-d_{u_2} \right|$ are the relative antenna element spacings of transmit and receive antenna arrays, respectively. $\sigma_{h_{u_1,s_1,n}}=\left | h_{u_1,s_1,n}(t) \right |$ and $\sigma_{h_{u_2,s_2,n}}=\left | h^*_{u_2,s_2,n}(t+\Delta t) \right |$ are the standard deviations of $h_{u_1,s_1,n}(t)$ and $h_{u_2,s_2,n}(t+\Delta t)$.
In order to demonstrate more clearly, the antenna polarization and antenna radiation pattern are not considered in the simulation. Uniform linear arrays (ULAs) are used at both ends and placed parallel to the $y$ axis.
By imposing $\Delta t=0$ and substituting (\ref{equ_CIR}) into (\ref{equ_spa_tem}), we get the spatial CCF between $h_{u_1,s_1,n}$ and $h_{u_2,s_2,n}$ as
\begin{align}
\rho_{(s_1, u_1), (s_2, u_2), n}(t,\Delta d_s, \Delta d_u)=& \rho_{(s_1, u_1), (s_2, u_2), n}^{\text{LoS}}(t,\Delta d_s,\Delta d_u)\nonumber \\ +&\rho_{(s_1, u_1), (s_2, u_2), n}^{\text{NLoS}}(t,\Delta d_s,\Delta d_u)
\label{equ_CCF}
\end{align}
where
\begin{align}
\rho&_{(s_1, u_1), (s_2, u_2), n}^{\text{LoS}}(t,\Delta d_s,\Delta d_u)=\frac{K \delta(n-1)}{K+1}\nonumber \\ &\cdot e^{jk[\Delta d_u \textup{cos} \gamma_\text{LoS}(t)\cdot \textup{sin}\varphi_\text{LoS}(t)+ \Delta d_s \textup{cos} \psi_\text{LoS}(t)\cdot \textup{sin}\phi_\text{LoS}(t)]}
\end{align}
and
\begin{align}
\rho&_{(s_1, u_1), (s_2, u_2), n}^{\text{NLoS}}(t,\Delta d_s,\Delta d_u)=\frac{1}{(K+1)M}\nonumber \\ &\sum_{m=1}^{M}\textup E\{ e^{jk[\Delta d_u \textup{cos} \gamma_{n,m}(t)\cdot \textup{sin}\varphi_{n,m}(t)+\Delta d_s \textup{cos} \psi_{n,m}(t)\cdot \textup{sin}\phi_{n,m}(t)]}\}.
\end{align}
Note that $\textup E\{ e^{\Phi_{n,m_1}-\Phi_{n,m_2}} \}=0$  when $m_1\neq m_2$. Equation (\ref{equ_CCF}) means the spatial CCF of the proposed non-stationary channel model not only depends on the relative antenna element spacings but also the time $t$. If we let $\Delta d_s=0$, which means two links share the same transmit antenna element, the corresponding spatial CCF can be calculated as
\begin{align}
\rho_{u_1,u_2,n}^\textup{MS}&(t,\Delta d_u)=\frac{K \delta(n-1)}{K+1}\cdot e^{jk\Delta d_u \textup{cos} \gamma_\text{LoS}(t)\cdot \textup{sin}\varphi_\text{LoS}(t)}\nonumber \\+&\frac{1}{(K+1)M}\sum_{m=1}^{M}\textup E\{ e^{jk\Delta d_u \textup{cos} \gamma_{n,m}(t)\cdot \textup{sin}\varphi_{n,m}(t)}\}.
\end{align}

\subsection{Temporal ACF}
The temporal ACF can be obtained by substituting (\ref{equ_CIR}) into (\ref{equ_spa_tem}) and imposing $\Delta d_u=0$ and $\Delta d_s=0$. Considering the non-stationarity of the channel model, a cluster has a probability of $e^{\frac{-\lambda_R}{D_c}\cdot(v_\textup{MS}\cdot \Delta t+P_c\cdot(v^A+v^Z)\cdot\Delta t)}$ to survive from $t$ to $t+\Delta t$. Therefore, we get the temporal ACF of the proposed channel model as
\begin{align}
r_n(t,\Delta t)= e^{\frac{-\lambda_R}{D_c} \cdot (v_\textup{MS} \cdot \Delta t+P_c\cdot(v_A+v_Z)\cdot\Delta t)}\nonumber \\ \cdot [r_n^\text{LoS}(t,\Delta t)+r_n^\text{NLoS}(t,\Delta t)]
\label{equ_ACF}
\end{align}
where
\begin{equation}
r_n^\text{LoS}(t,\Delta t)=\frac{K\delta(n-1)}{K+1}\cdot e^{j2\pi[\cdot \nu_\text{LoS}(t)\cdot t-\nu_\text{LoS}(t+\Delta t)\cdot (t+\Delta t)]}
\end{equation}
and
\begin{align}
r_n^\text{NLoS}&(t,\Delta t)=\frac{1}{(K+1)M} \nonumber \\ &\sum_{m=1}^{M}\text E\{ e^{j2\pi\cdot [\nu_{n,m}(t)\cdot t-\nu_{n,m}(t+\Delta t)\cdot (t+\Delta t)]}\}
\label{equ_ACF_NLoS}.
\end{align}
Equation (\ref{equ_ACF}) indicates that the temporal ACF depends on both interval $\Delta t$ and time $t$.

\subsection{Doppler PSD}
The Doppler PSD $S_n(f,t)$ of the proposed channel model is obtained by the Fourier transform of the temporal ACF $r_n(t,\Delta t)$ with respect to the time interval $\Delta t$, and can be written as
\begin{equation}
S_n(f,t)=\int_{-\infty }^{\infty}r_n(t,\Delta t)e^{-j2\pi f\Delta t}d(\Delta t).
\end{equation}
Note that the Doppler PSD of the non-stationary channel model is time dependent.
\subsection{Time-Variant Transfer Function}
The time-variant transfer function of the channel $H_{u,s}(f,t)$ is defined as the Fourier transform of the time-variant impulse response $h_{u,s}(t,\tau)$  with respect to the propagation delay $\tau$, and is obtained as
\begin{align}
H_{u,s}(f,t)&=\int_{-\infty }^{\infty}h_{u,s}(t,\tau)e^{-j2\pi f\tau}d\tau \nonumber\\
                    &=\sum_{n=1}^{N(t)}h_{u,s,n}(t)e^{-j2\pi f\tau_n(t)}.
\end{align}
\subsection{ LCR and  AFD}
The LCR and AFD provide information about how fast the channel changes with time, which are important statistical properties for the design and performance analysis of wireless communication system. The LCR is defined as average number of times the envelope of signal crosses a given threshold from up to down (or from down to up), and can be calculated as \cite{Patzold_LCR}
\begin{align}
N(r,t) = &{2r\sqrt{K + 1} \over \pi^{3/2}}\sqrt{{b_{2}(t) \over b_{0}} - {b_{1}^{2}(t)\over b_{0}^{2}}}\ e^{ - K - (K+1)r^{2}}  \nonumber\\ &\cdot \int_{0}^{\pi/2}\cosh \ \Big(2\sqrt{K(K + 1)}r\cos\theta \Big) \nonumber \\ &\cdot \Big[e^{-(\chi(t) \sin \theta)^{2}} + \sqrt{\pi}\chi(t) \sin \theta \ {\rm erf}\ (\chi(t)\sin\theta) \Big]\ d\theta
\end{align}
where $\cosh(\cdot)$ is the hyperbolic cosine function and $\rm erf(\cdot)$ is the error function. Parameters $\chi(t)$ and $b_l(t) (l=0,1,2)$ are calculated as
\begin{equation}
\chi(t)=\sqrt\frac{K\cdot b_1(t)^2}{b_0 \cdot b_2(t)-b_1(t)^2}
\end{equation}
and
\begin{equation}
b_l(t)=\frac{d^l r^\text{NLoS}_n(t,\Delta t)}{j^l d \Delta t^l}.
\end{equation}
By using (\ref{equ_ACF_NLoS}), the $b_l(t)$ can be can be further expressed as
\begin{align}
b_l(t)=&\frac{(2\pi)^l}{(K+1)\lambda^lM} \cdot\sum_{m=1}^{M}(\overline v_{\textup{MS}}\cdot\overline\Psi_{m}(t)\nonumber\\&-\overline v_n^A\cdot\overline\Phi_{m}(t)-\overline v_n^Z\cdot\overline\Psi_{m}(t))^l.
\end{align}
Note that the calculation of LCR is based on a narrowband channel, and the first path is used in the calculation, i.e., $\overline\Psi_m(t)=\overline\Psi_{1,m}(t)$ and $\overline\Phi_m(t)=\overline\Phi_{1,m}(t)$.
It is worth mentioning that key parameters $b_1(t)$ and $b_2(t)$ are time-varying and can capture the effects of both the MS and cluster movements, which is different from the expressions in \cite{Patzold_LCR,Alenka_LCR}.

AFD is defined as average time duration the envelope remains below a given threshold. It is defined as \cite{Patzold_LCR}
\begin{equation}
 L(r,t) = {{1 - {\rm{Q}}\left({\sqrt {2K}, \sqrt {2(K + 1){r^2}} } \right)} \over { N(r,t)}}
\end{equation}
where $Q(\cdot,\cdot)$ denotes the Marcum $Q$ function.

\subsection{Stationary Interval}
 The stationary interval is the maximum time duration over which the
WSS assumption is valid and can be calculated through the local region of stationarity (LRS) \cite{StationarityInterval} that defines the stationary region with respect to the channel power. The LRS can be calculated by the largest interval within which the correlation coefficient between two consecutive averaged power delay profiles (APDPs) beyond a certain threshold. The correlation coefficient is defined as
 \begin{equation}
 c(t_k,\Delta t)=\frac{\int \overline{P_h}(t_k,\tau)\cdot \overline{P_h}(t_k+\Delta t,\tau)d\tau}{\textup{max}\{  \int \overline{P_h}(t_k,\tau)^2d\tau, \overline{P_h}(t_k+\Delta t,\tau)^2 d\tau \}}
 \end{equation}
where  $\overline{P_h}(t_k,\tau)$ represents the APDP and can be expressed as
\begin{equation}
\overline{P_h}(t_k,\tau)=\frac{1}{N_\textup{PDP}}\sum_{k}^{k+N_\textup{PDP}-1}\left| h_{u,s}(t_k,\tau) \right|^2
\end{equation}
where $N_\textup{PDP}$ is the number of power delay profiles to be averaged, $k$ represents temporal (drop) index, and $ h_{u,s}(t_k,\tau) =\sum_{n=1}^{N(t_k)} h_{u,s,n}(t_k)\delta(\tau-\tau_n)$. The stationarity interval $T_s(t_k)$  is the maximum time duration over which LRS correlation coefficient beyond a given threshold, i.e., $c_\textup{thresh}$, and can be calculated as
\begin{equation}
T_s(t_k)=\textup{max}\{\Delta t\mid c(t_k,\Delta t)\geqslant c_\textup{thresh}\}.
\end{equation}

\section{RESULTS AND ANALYSIS}
\label{simulation_results_and_analysis}
\begin{figure}
\centering\includegraphics[width=3.5in]{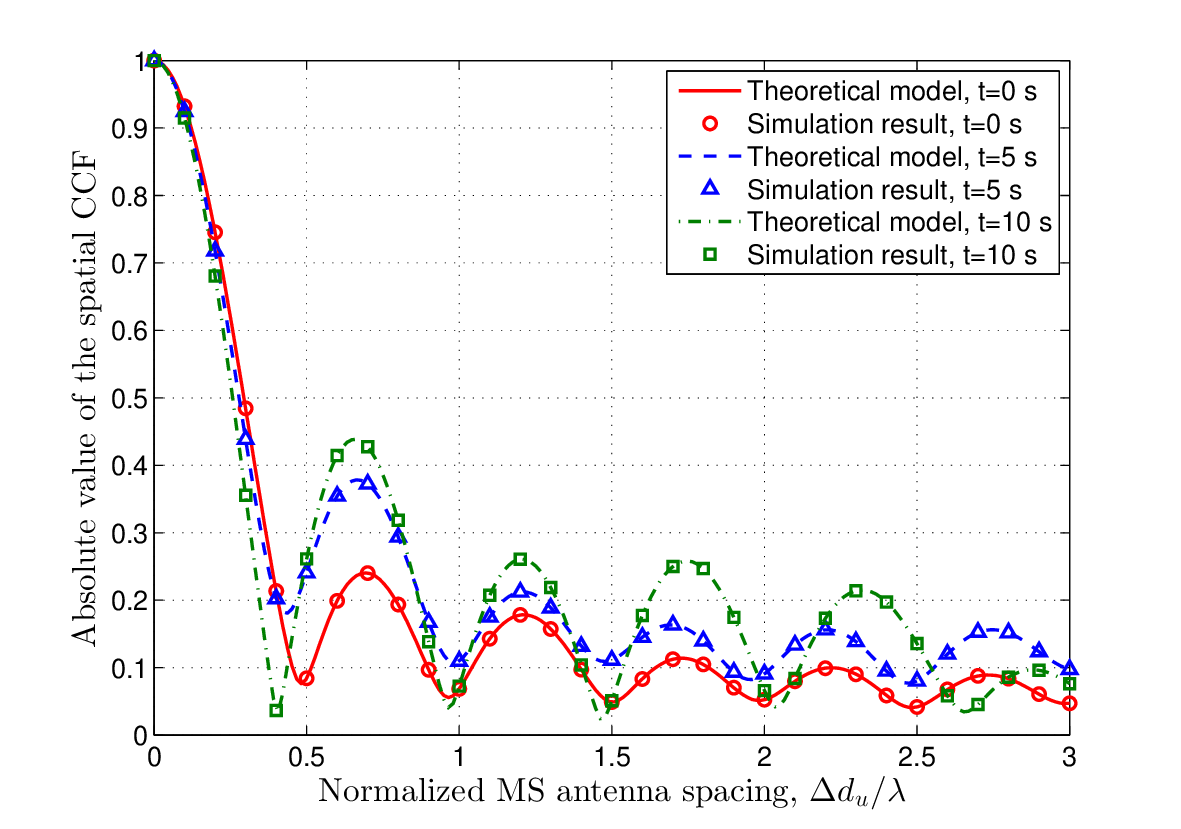}
\centering\caption{The theoretical and simulated spatial CCFs of the 3-D non-stationary channel model at different time instants (UMi NLoS scenario, $D^T_n(t_0) =200 $ m, $D^R_n(t_0) =200 $ m, $v_{\textup{MS}}=20$ m/s, $\theta_{\textup{MS}}=120^{\circ}$, $\vartheta_{\textup{MS}}=0^{\circ}$, $v^A_n$ and $v^Z_n \sim \textup U(0,10)$ m/s, $\theta^A_n$ and $\theta^Z_n \sim \textup U(-180^{\circ},180^{\circ})$, $\vartheta^A_n$ and $\vartheta^Z_n \sim  \textup U(-90^{\circ},90^{\circ})$). }
\label{fig_CCF_2D}
\end{figure}
The absolute values of the local spatial CCFs of the 3-D non-stationary channel model in UMi scenario with NLoS condition at three time instants are illustrated in Fig. \ref{fig_CCF_2D} (the curve of ``$t=0$" was obtained from the original WINNER+ channel model). The antenna spacing is normalized with respect to the wavelength. The absolute values of the spatial CCFs vary with time $t$ due to the drafting of angle parameters. A good consistent between theoretical results and simulated results ensures the correctness of our simulations and derivations.

\begin{figure}
\centering\includegraphics[width=3.5in]{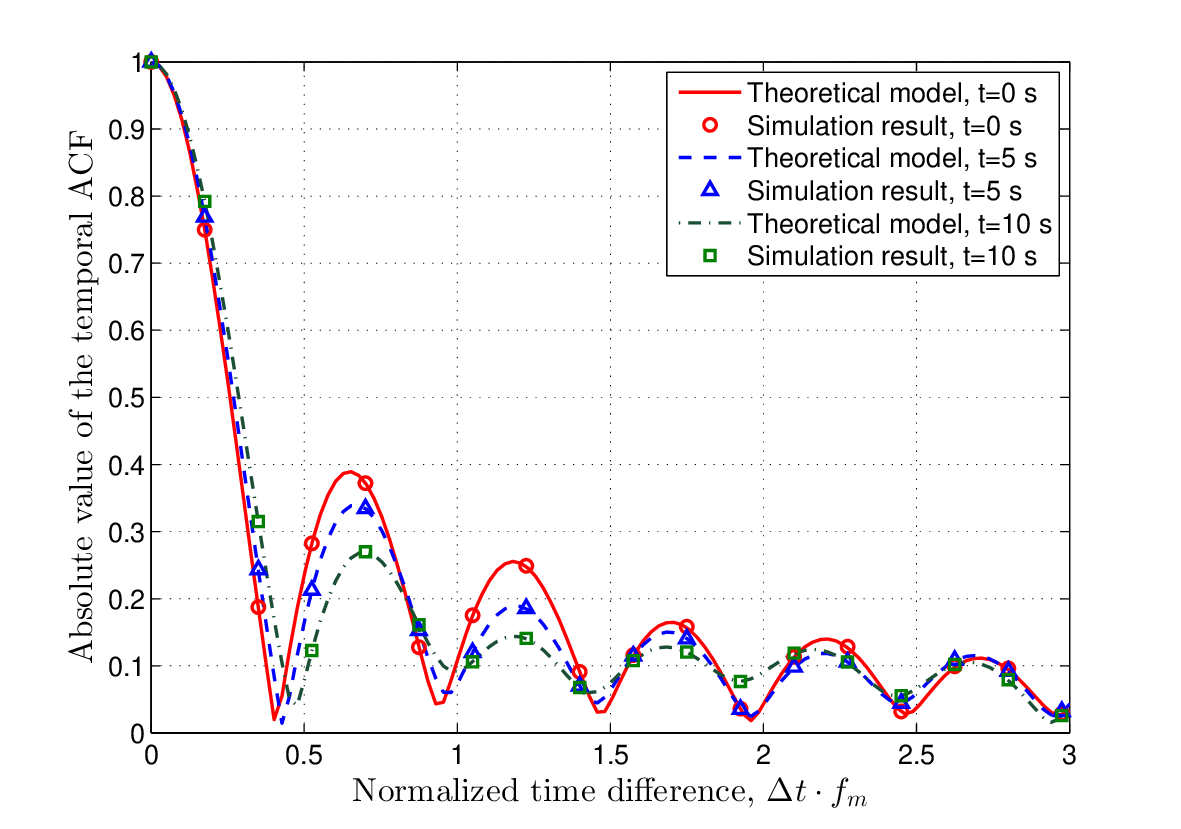}
\centering\caption{The theoretical and simulated temporal ACFs of the 3-D non-stationary channel model at different time instants (UMi NLoS scenario, $f_m=\left| v_{\textup{MS}} \right| / \lambda$, $D^T_n(t_0) =200 $ m, $D^R_n(t_0) =200 $ m, $v_{\textup{MS}}=20$ m/s, $\theta_{\textup{MS}}=120^{\circ}$, $\vartheta_{\textup{MS}}=0^{\circ}$,  $v^A_n$ and $v^Z_n \sim \textup U(0,10)$ m/s, $\theta^A_n$ and $\theta^Z_n \sim \textup U(-180^{\circ},180^{\circ})$, $\vartheta^A_n$ and $\vartheta^Z_n \sim  \textup U(-90^{\circ},90^{\circ})$).  }
\label{fig_ACF_2D}
\end{figure}
The absolute values of the temporal ACFs of the 3-D non-stationary channel model in the UMi scenario with NLoS condition at three different time instants are shown in Fig. \ref{fig_ACF_2D}. The time difference is normalized with respect to the wavelength. The shifting of the ACF is mainly resulted from the Doppler shifts caused by the movements of the MS and clusters. In addition, the figure shows the simulation results match the theoretical results very well, verifying the correctness of the simulations and derivations.

\begin{figure}
\centering\includegraphics[width=3.5in]{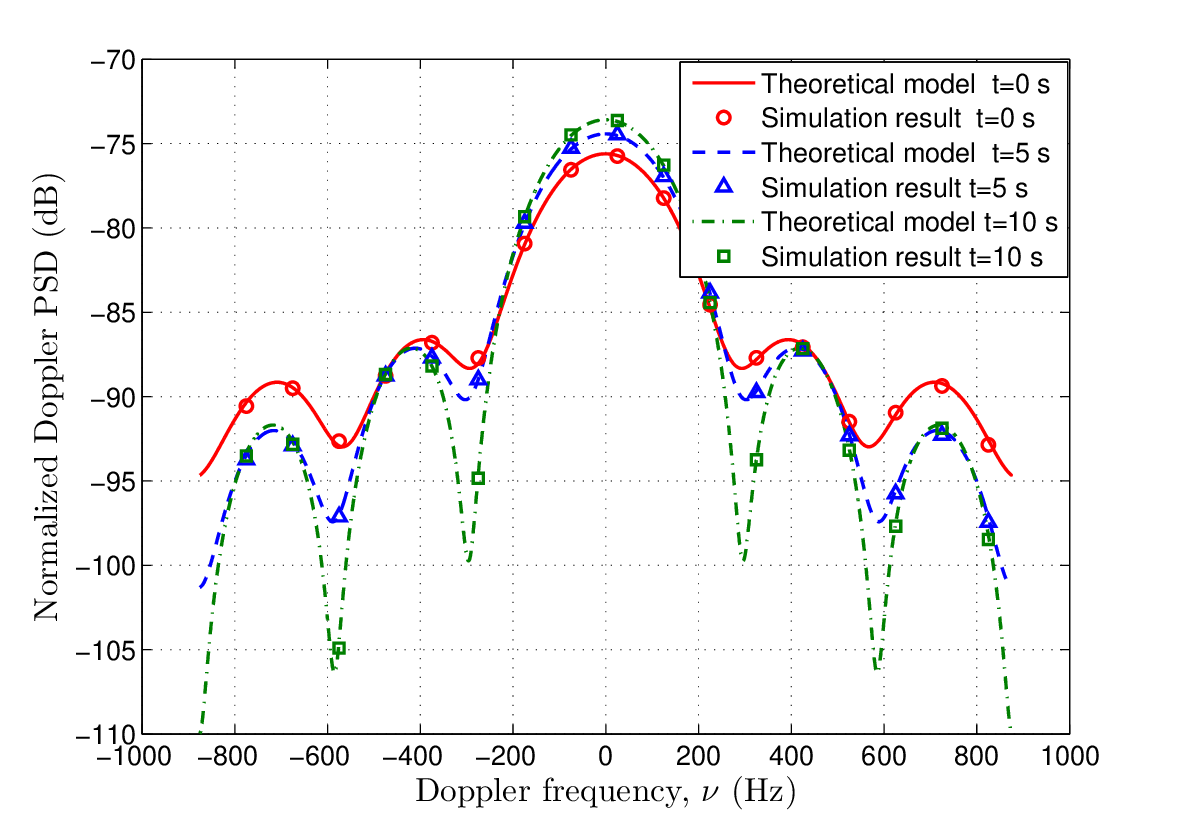}
\centering\caption{The theoretical and simulated normalized Doppler PSDs at different time instants (UMa NLoS scenario, $D^T_n(t_0) = 100$ m, $D^R_n(t_0) = 100$ m, $v_{\textup{MS}}=50$ m/s, $\theta_{\textup{MS}}=60^{\circ}$, $\vartheta_{\textup{MS}}=0^{\circ}$, $v^A_n$ and $v^Z_n \sim \textup U(0,20)$ m/s, $\theta^A_n$ and $\theta^Z_n \sim \textup U(-180^{\circ},180^{\circ})$), $\vartheta^A_n$ and $\vartheta^Z_n \sim \textup U(-90^{\circ},90^{\circ})$). }
\label{fig_doppler}
\end{figure}

The Doppler PSDs of the proposed channel model at three different time instants are illustrated in Fig. \ref{fig_doppler}. The Doppler PSDs  drift over time due to the movements of MS and clusters can be clearly observed. Again, the simulation results align well with the theoretical results, illustrating the correctness of the
simulations and derivations.
\begin{figure}
\begin{minipage}[t]{0.5\linewidth}
\centering
\includegraphics[width=1.8in]{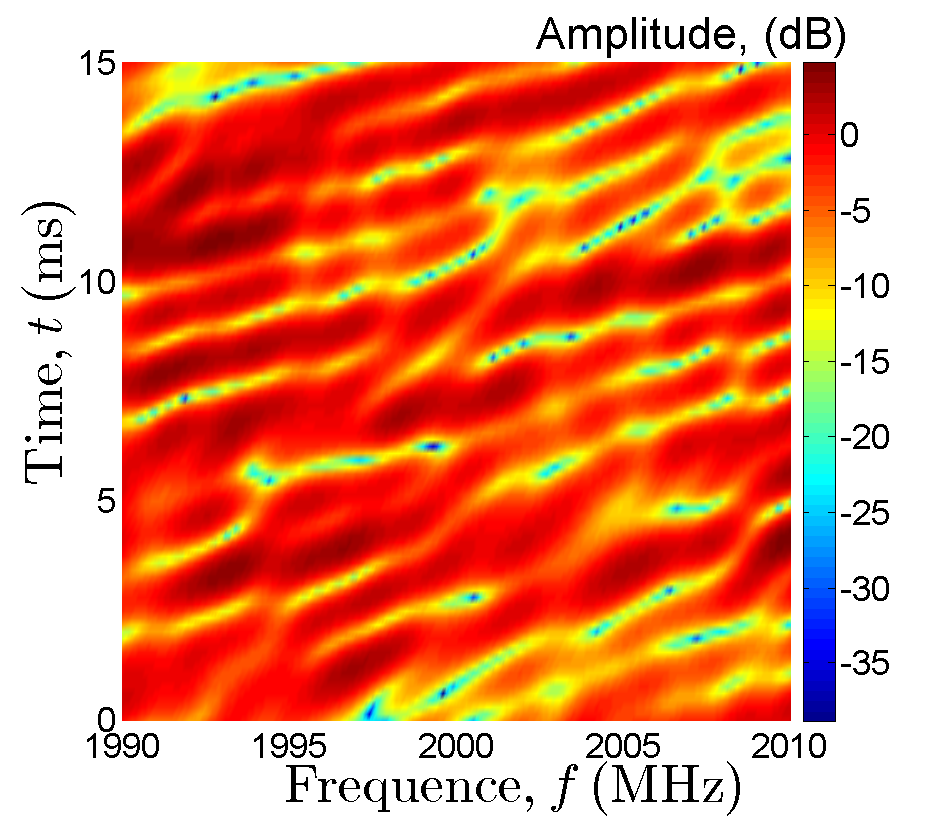}
\centerline {(a)} \tiny
\end{minipage}%
\begin{minipage}[t]{0.5\linewidth}
\centering
\includegraphics[width=1.8in]{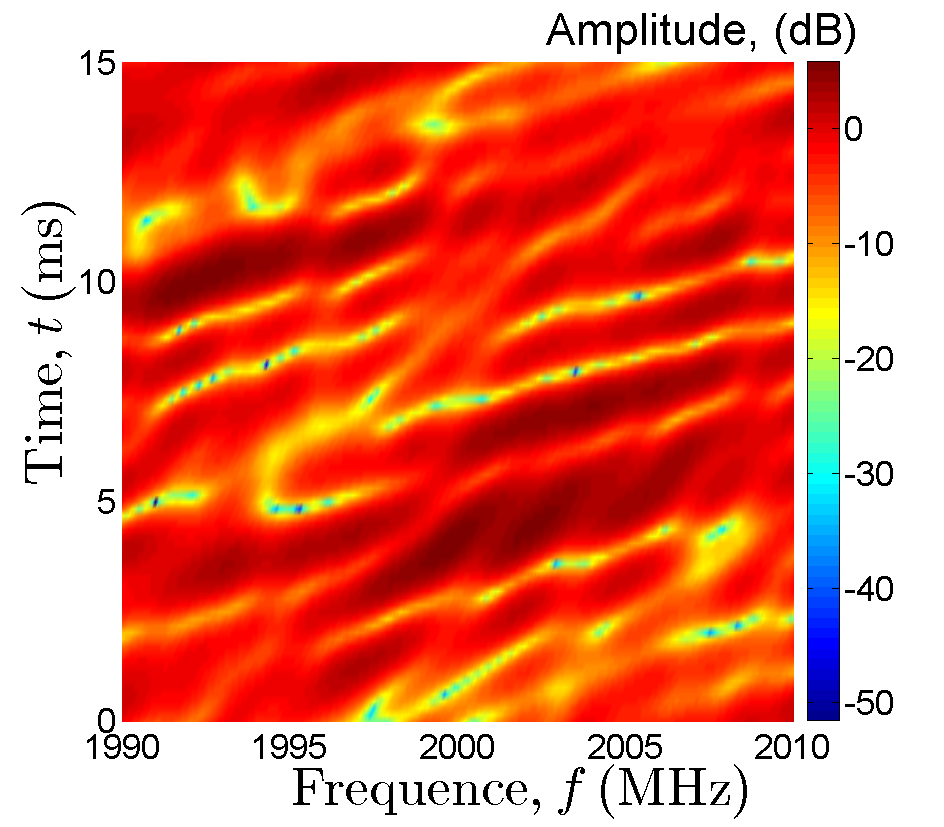}
\centerline{(b)}\tiny
\end{minipage}
\caption{Time-variant transfer functions of (a) the 3-D
non-stationary channel model and (b)  the WINNER+ channel model (UMi NLoS scenario, $f_c=2$ GHz, SampleDensity = 16, $D^T_n(t_0) = 100$ m, $D^R_n(t_0) = 100$ m, $v_{\textup{MS}}=60$ m/s, $\theta_{\textup{MS}}=60^{\circ}$, $\vartheta_{\textup{MS}}=0^{\circ}$, $v^A_n$ and $v^Z_n \sim \textup U(0,20)$ m/s, $\theta^A_n$ and $\theta^Z_n \sim \textup U(-180^{\circ},180^{\circ})$), $\vartheta^A_n$ and $\vartheta^Z_n \sim \textup U(-90^{\circ},90^{\circ})$).}
\label{fig_FreRes_WSS}
\end{figure}


The time-variant transfer functions  of the 3-D non-stationary channel model (left-side) and the WINNER+ channel model (right-side) are illustrated in Fig. \ref{fig_FreRes_WSS}.  In order to make a clear comparison, the time-variant transfer functions are illustrated in 2-D. The band of interest is 1990 MHz to 2010 MHz with a sample spacing 0.1 MHz. The time-variant transfer functions of proposed channel model contain more fluctuations on both time and frequency axes than the ones of the WINNER+ channel model due to its non-stationary properties, which include more realistic features and details than the case in WINNER+ channel model.
\begin{figure}
\centering\includegraphics[width=3.5in]{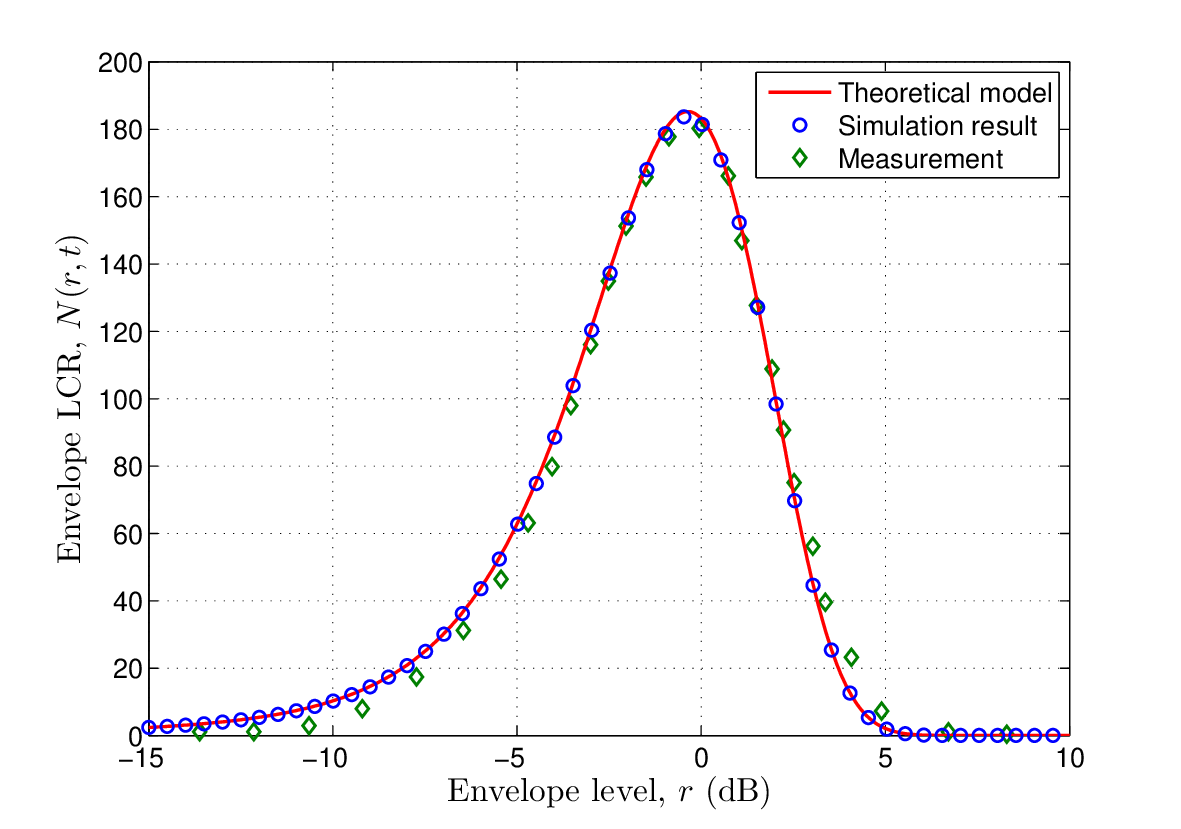}
\centering\caption{The theoretical and simulated envelope LCRs for the 3-D non-stationary channel model vs. measurement data in \cite{lcrafd_measurement} (UMa LoS scenario, $f_c=930.2$ MHz, $K$ = 5.514, $D^T_n(t_0) = 200$ m, $D^R_n(t_0) = 200$ m, $v_{\textup{MS}}=90$ m/s, $\theta_{\textup{MS}}=90^{\circ}$, $\vartheta_{\textup{MS}}=0^{\circ}$, $v^A_n$ and $v^Z_n \sim \textup U(0,5)$ m/s, $\theta^A_n$ and $\theta^Z_n \sim \textup U(-180^{\circ},180^{\circ})$), $\vartheta^A_n$ and $\vartheta^Z_n \sim \textup U(-90^{\circ},90^{\circ})$).}
\label{fig_lcr}
\end{figure}

\begin{figure}
\centering\includegraphics[width=3.5in]{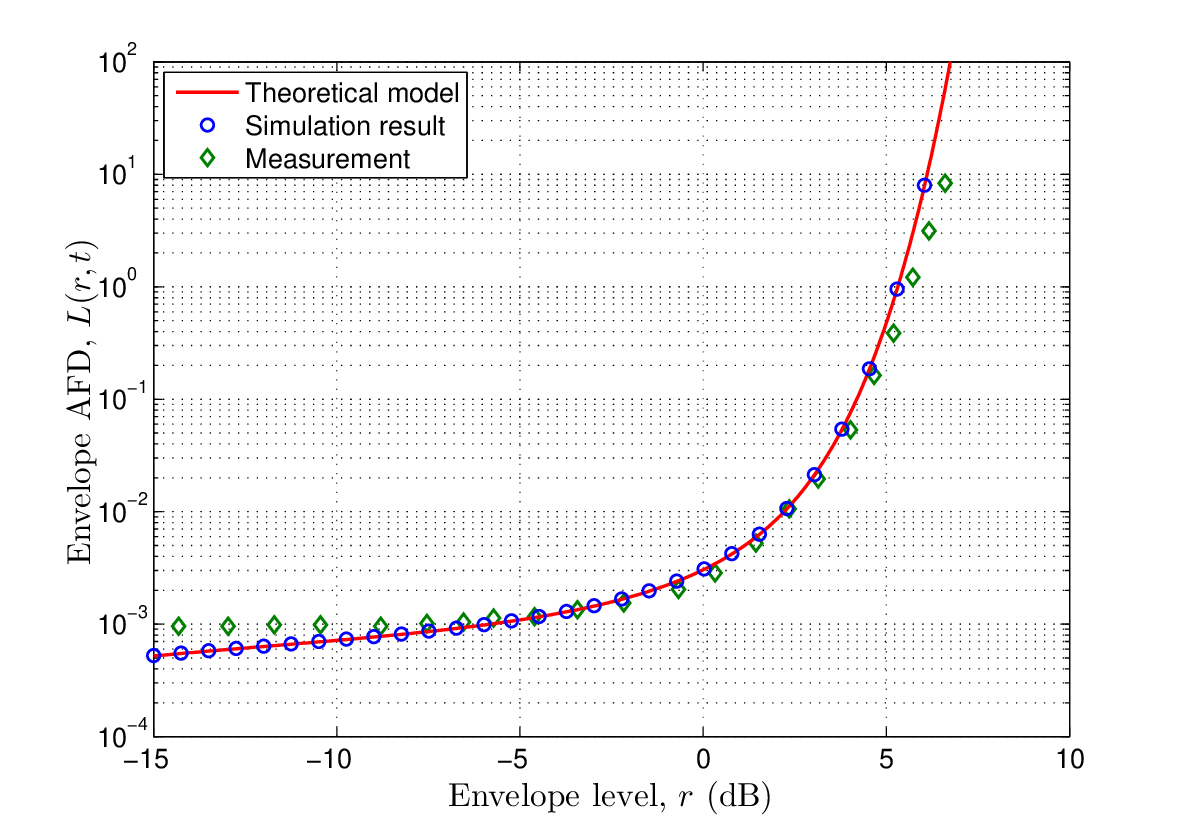}
\centering\caption{The theoretical and simulated envelope AFDs for the 3-D non-stationary channel model vs. measurement data in \cite{lcrafd_measurement}. (UMa LoS scenario, $f_c=930.2$ MHz, $K$ = 5.514, $D^T_n(t_0) = 200$ m, $D^R_n(t_0) = 200$ m, $v_{\textup{MS}}=90$ m/s, $\theta_{\textup{MS}}=90^{\circ}$, $\vartheta_{\textup{MS}}=0^{\circ}$, $v^A_n$ and $v^Z_n \sim \textup U(0,5)$ m/s, $\theta^A_n$ and $\theta^Z_n \sim \textup U(-180^{\circ},180^{\circ})$), $\vartheta^A_n$ and $\vartheta^Z_n \sim \textup U(-90^{\circ},90^{\circ})$).}
\label{fig_afd}
\end{figure}

Fig. \ref{fig_lcr} shows the theoretical and simulation envelope LCRs for the 3-D non-stationary channel model vs. measurement data obtained in \cite{lcrafd_measurement}. The measurement was conducted in HST scenario at 930.2 MHz with a speed of 300 km/h. The simulation results are calculated by counting the number of times the envelope crosses different threshold levels and being divided by the observation period. Meanwhile,  a good consistent between the simulation results and the theoretical results can be observed, both of which are validated by the measurement data.

Fig. \ref{fig_afd} shows the theoretical and simulation envelope AFDs for the 3-D non-stationary channel model vs. measurement data in \cite{lcrafd_measurement}. The simulation results are calculated by keeping track of all fade durations and being divided by the total number of fades. Again, a good agreement between the simulation results and theoretical results can be observed. Furthermore, the simulation and the theoretical results are compared and validated by the measurement data.

\begin{figure}
\centering\includegraphics[width=3.5in]{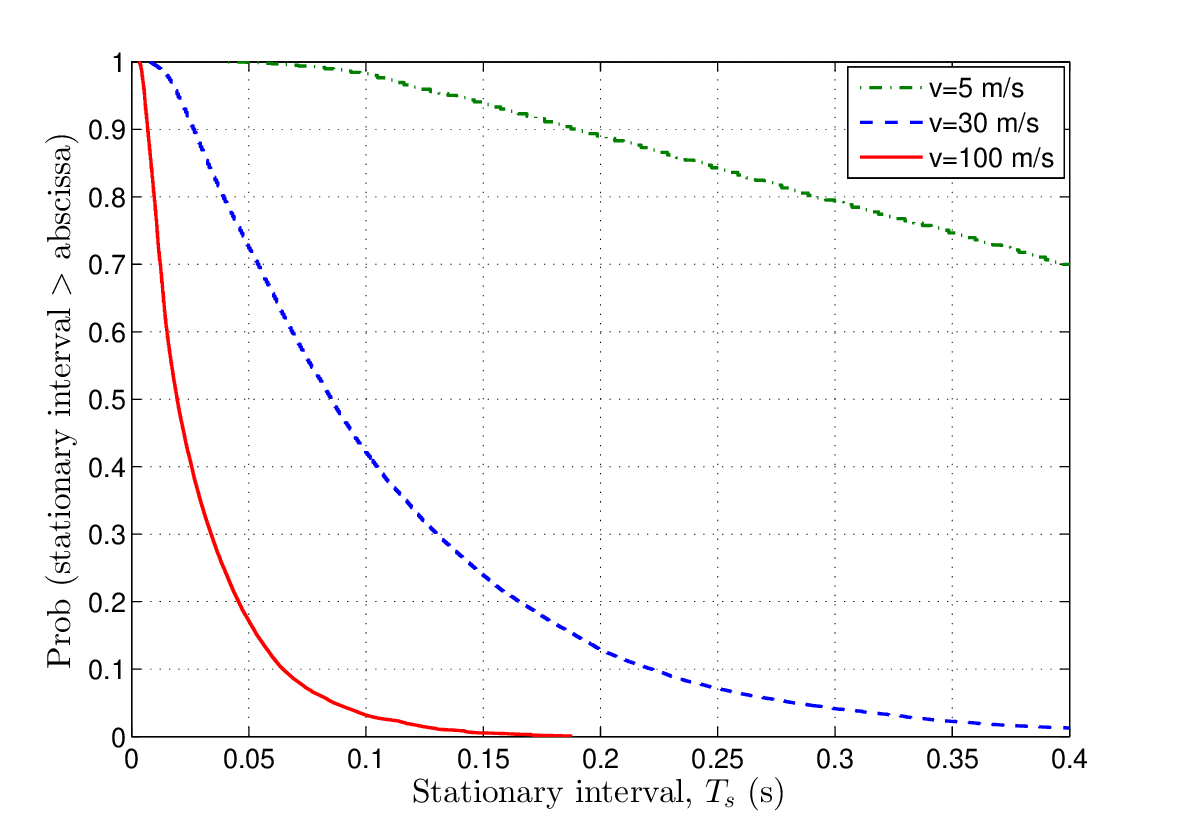}
\centering\caption{The CCDFs of stationary intervals for the 3-D non-stationary channel model at three different speeds (UMa LoS scenario, $f_c=930.2$ MHz, SampleDensity = 4, $c_{\textup{thresh}}=0.8$, $D^T_n(t_0) =100 $ m, $D^R_n(t_0) =70 $ m, $\theta_{\textup{MS}}=120^{\circ}$, $\vartheta_{\textup{MS}}=0^{\circ}$, $v^A_n$ and $v^Z_n \sim \textup U(0,20)$ m/s, $\theta^A_n$ and $\theta^Z_n \sim \textup U(-180^{\circ},180^{\circ})$, $\vartheta^A_n$ and $\vartheta^Z_n \sim  \textup U(-90^{\circ},90^{\circ})$). }
\label{fig_CCDF_SI_V}
\end{figure}

\begin{figure}
\centering\includegraphics[width=3.5in]{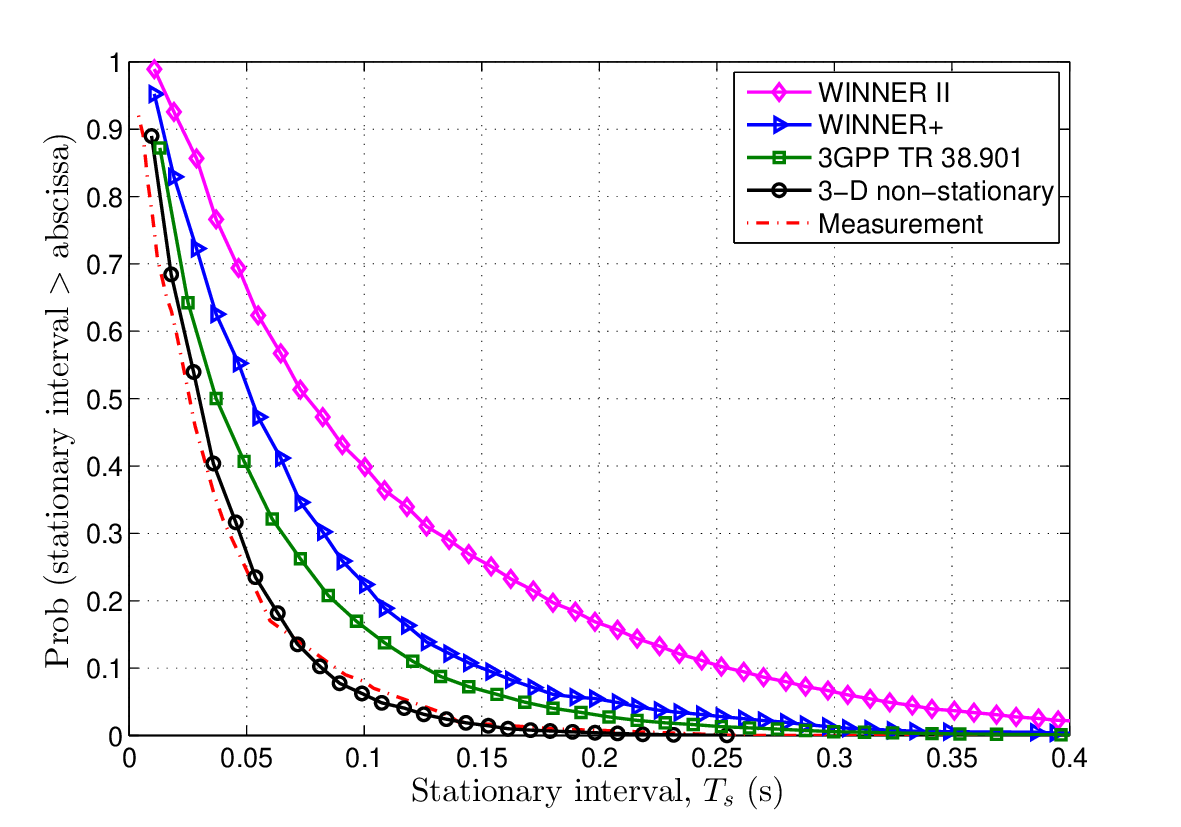}
\centering\caption{The CCDFs of stationary intervals for the WINNER II, WINNER+, 3GPP TR 38.901, 3-D non-stationary channel model, and the measured channel in \cite{StationarityInterval} (UMa LoS scenario, $f_c=930.2$ MHz, SampleDensity = 4, $c_{\textup{thresh}}=0.8$, $D^T_n(t_0) =100 $ m, $D^R_n(t_0) =70 $ m, $v_\textup{MS}=90$ m/s, $\theta_{\textup{MS}}=-60^{\circ}$, $\vartheta_{\textup{MS}}=0^{\circ}$, $v^A_n =0$ m/s, $v^Z_n=30$ m/s, $\theta^Z_n=120^{\circ}$, and $\vartheta^Z_n=0^{\circ}$). }
\label{fig_CCDF_SI}
\end{figure}

The empirical complementary cumulative distribution functions (CCDFs) of the stationary intervals for the proposed 3-D non-stationary channel model are shown in Fig. \ref{fig_CCDF_SI_V}. The simulations are conducted at three different speeds, i.e., 5~m/s, 30 m/s, and 100 m/s.  The abscissa values of this figure are stationary intervals in seconds and the ordinate values are probability of the stationary intervals larger than the abscissa values. It can be observed that at speed 100 m/s, in 80\% of the case, the channel can be considered as stationary over a time interval of 9.5 ms, 39 ms at speed 30 m/s, and 292 ms at 5 m/s, i.e., the higher the speed is, the shorter the stationary interval will be, which is consistent with the theoretical analysis and the measurement results. Besides the speed of MS, other  factors such as movement direction of MS, movement directions of clusters, and speeds of clusters also impact on the results as well.

The empirical CCDFs of stationary intervals for the proposed 3-D non-stationary channel model,  the measured HST channel in\cite{StationarityInterval}, and three existing channel models, i.e., WINNER II, WINNER+, and 3GPP TR 38.901 \cite{3GPP38901} are shown in Fig. \ref{fig_CCDF_SI}.
The measurement data was obtained from channel measurement conducted in Zhengxi passenger dedicated line with a speed of 200 km/h $\sim$ 340 km/h. The speed of MS in the proposed channel model and the three existing channel model are set to 90 m/s, which is almost the same as the maximum speed in the measurement campaign.
The stationary interval of the proposed channel model is equal to 11 ms in 80\% of the case and 21 ms in 60\% of the case, which is almost the same as the measurement data, i.e., 9 ms in 80\% of the case and 20 ms in 60\% of the case.
However, the three widely used channel models illustrate larger stationary intervals than that of the proposed channel model and the measurement data. For the first two models, i.e., WINNER II and WINNER+, the time evolution and cluster movement are not considered, and the channel variance over time mainly results from the phase changes of the multipath and the Doppler shift caused by the MS movement.
The latest 3GPP channel model, i.e., 3GPP TR 38.901, took into account time-varying parameters (generated using ``Procedure A" \cite{3GPP38901} in the simulation), which results in the shortest stationary interval among the three existing channel models. However, the 3GPP TR 38.901 channel model still did not consider cluster movement, which may overlook certain non-stationary properties of fast changing channels.

\section{Conclusions} \label{conclusion_section}
A novel 3-D non-stationary wideband MIMO channel model based on WINNER+ channel model has been proposed in this paper. The proposed channel model has taken into account both fixed clusters and moving clusters. Transition regions have been used to provide a smooth time evolution. Time-variant parameters such as positions of clusters, angle parameters, delays, and path powers have been derived using geometric methods based on the movements of the MS and clusters. Statistical properties including spatial CCF, temporal ACF, Doppler PSD,  LCR, AFD, and stationary interval have been investigated. The simulation results match the theoretical results very well, showing the correctness of both derivations and simulations. The LCR, AFD, and stationary interval of the proposed model have been validated by the measurement data which illustrates the usefulness of the proposed channel model. Furthermore, the stationary interval of the proposed model has been compared with those of different existing channel models, showing the capability of the proposed model capturing the characteristics of non-stationary fast-changing channels.

Considering the fact that many existing millimetre wave channel models, e.g., METIS channel model \cite{METISD1.4}, 5GCM \cite{5GCM}, 3GPP TR 38.901 \cite{3GPP38901}, mmMAGIC channel model \cite{mmMagic} adopted the WINNER-like channel model frame work, the proposed channel model can be easily extended to support millimetre wave bands through a few additional modeling components. Specifically, intra-cluster delays and powers should be determined unequally and offset angles within a cluster should be generated randomly rather than using fixed values in order to achieve high temporal and spatial resolutions. The number of clusters and the number of rays within a cluster should be determined stochastically based on millimetre wave measurement results. Besides, the oxygen absorption and blockage modeling should also be considered.




%



%
%

\ifCLASSOPTIONcaptionsoff
  \newpage
\fi



%
%
\iftrue

\fi


\begin{IEEEbiography}[{\includegraphics[width=1.1in,height=1.3in,clip,keepaspectratio] {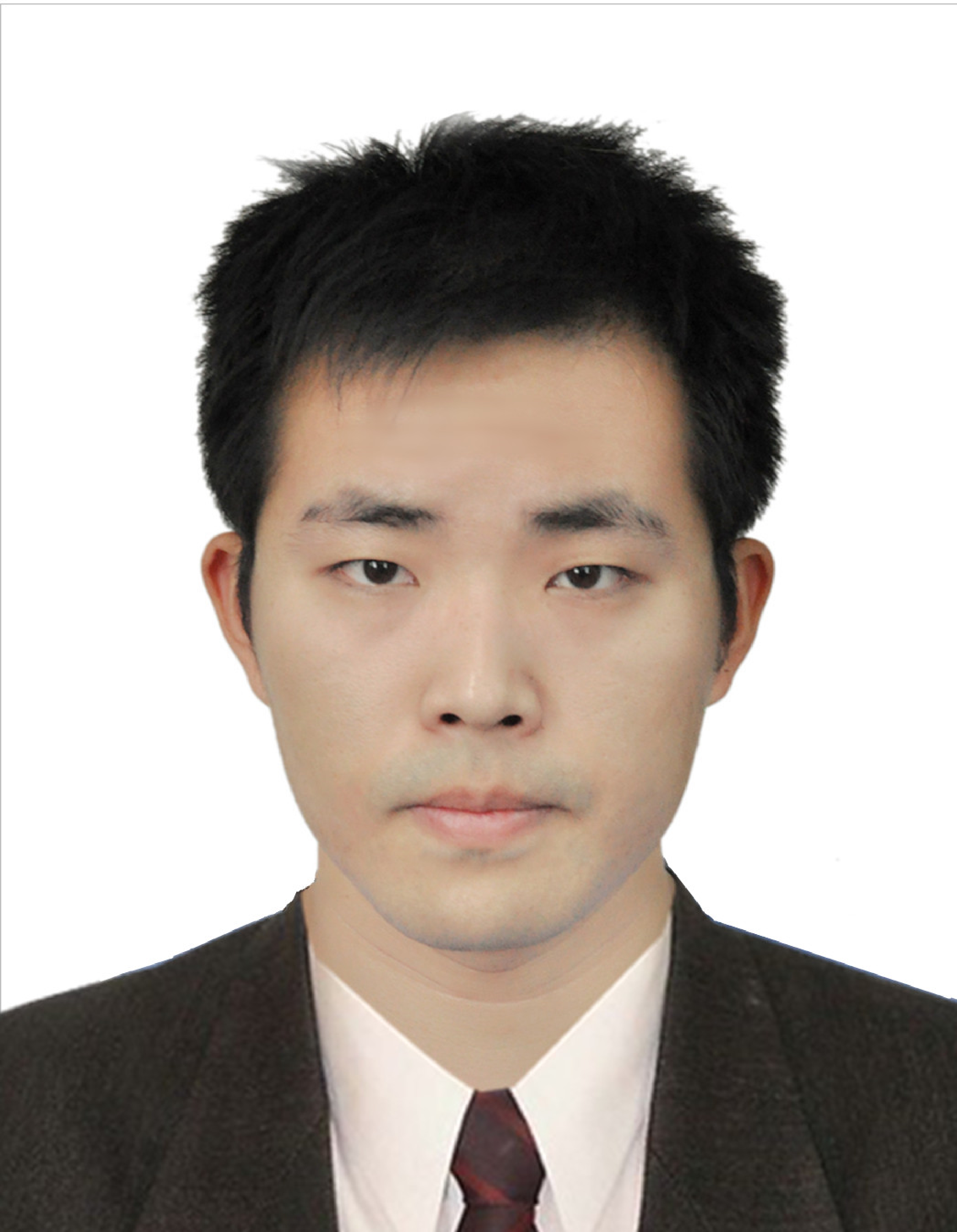}}]
{Ji Bian}received the B.Sc. degree in electronic information science and technology from Shandong Normal University, Jinan, Chian, in 2010 and the M.Sc. degree in signal and information processing from Nanjing University of Posts and Telecommunications, Nanjing, China, in 2013.

He is currently pursuing the Ph.D. degree at the School of Information Science and Engineering, Shandong University, Shandong, China. His current research interests include channel measurements, wireless propagation channel characterization, and 5G channel modeling.
\end{IEEEbiography}

\begin{IEEEbiography}[{\includegraphics[width=1.1in,height=1.3in,clip,keepaspectratio] {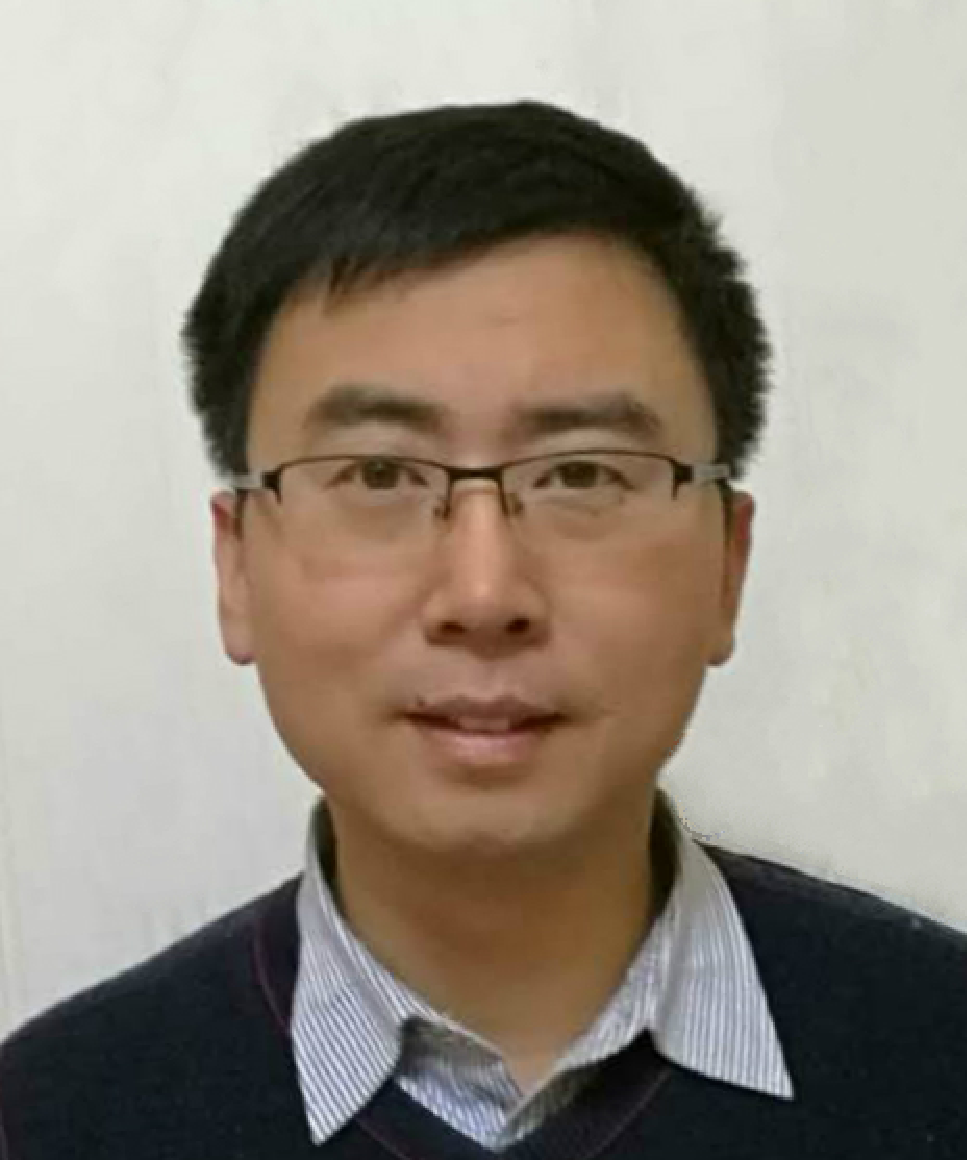}}]
{Jian Sun} Jian Sun (M'08) received the BSc in  Applied Electronic Technology, MEng in Measuring and Testing Technologies and Instruments, and PhD degrees in Communication and Information Systems, all from Zhejiang University, Hangzhou, China, in 1996, 1999 and 2005, respectively.

Since July 2005, he has been a Lecturer in the School of Information Science and Engineering, Shandong University, China. In 2011, he was a visiting scholar at Heriot-Watt University, UK, supported by UK-China Science Bridges: R\&D on (B)4G Wireless Mobile Communications (UC4G) project. His current research interests are in the areas of signal processing for wireless communications, channel sounding and modeling, propagation measurement and parameter extraction, MIMO and multicarrier transmission systems design and implementation.
\end{IEEEbiography}

\begin{IEEEbiography}[{\includegraphics[width=1.1in,height=1.3in,clip,keepaspectratio] {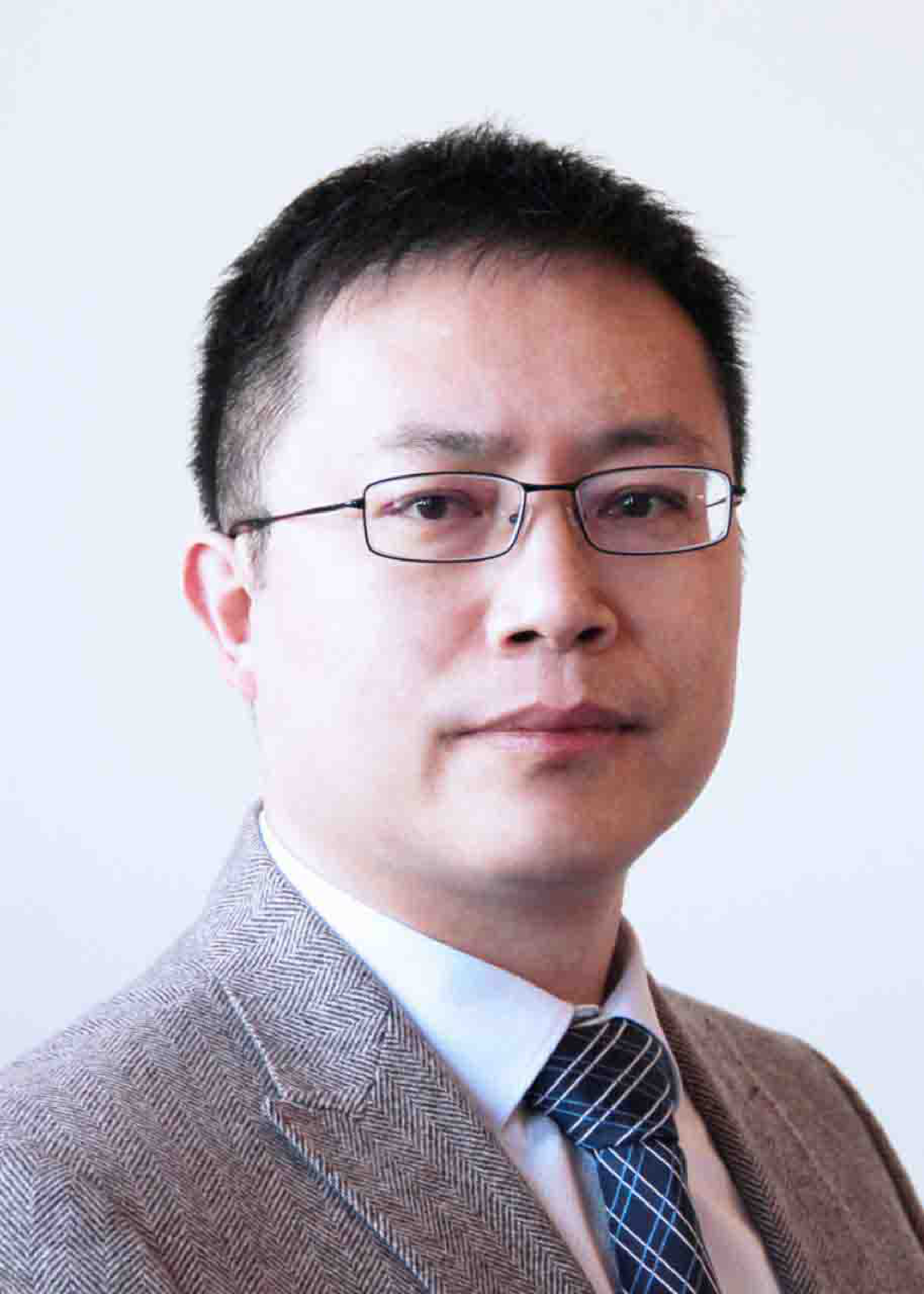}}]
{Cheng-Xiang Wang (S'01-M'05-SM'08-F'17)} received the BSc and MEng degrees in Communication and Information Systems from Shandong University, China, in 1997 and 2000, respectively, and the PhD degree in Wireless Communications from Aalborg University, Denmark, in 2004.

He was a Research Fellow with the University of Agder, Grimstad, Norway, from 2001 to 2005, a Visiting Researcher with Siemens AG Mobile Phones, Munich, Germany, in 2004, and a Research Assistant with the Hamburg University of Technology, Hamburg, Germany, from 2000 to 2001. He has been with Heriot-Watt University, Edinburgh, U.K., since 2005, where he was promoted to a Professor in 2011. He is also an Honorary Fellow of the University of Edinburgh, U.K., and a Chair/Guest Professor of Shandong University and Southeast University, China. He has authored 2 books, one book chapter, and over 310 papers in refereed journals and conference proceedings. His current research interests include wireless channel modeling and (B)5G wireless communication networks, including green communications, cognitive radio networks, high mobility communication networks, massive MIMO, millimetre wave communications, and visible light communications.

Prof. Wang is a Fellow of the IET and HEA, and a member of the EPSRC Peer Review College. He served or is currently serving as an Editor for nine international journals, including the IEEE TRANSACTIONS ON VEHICULAR TECHNOLOGY since 2011, the IEEE TRANSACTIONS ON COMMUNICATIONS since 2015, and the IEEE TRANSACTIONS ON WIRELESS COMMUNICATIONS from 2007 to 2009. He was the leading Guest Editor of the IEEE JOURNAL ON SELECTED AREAS IN COMMUNICATIONS, Special Issue on Vehicular Communications and Networks. He is also a Guest Editor of the IEEE JOURNAL ON SELECTED AREAS IN COMMUNICATIONS, Special Issue on Spectrum and Energy Efficient Design of Wireless Communication Networks and Special Issue on Airborne Communication Networks, and the IEEE TRANSACTIONS ON BIG DATA, Special Issue on Wireless Big Data. He served or is serving as a TPC Member, TPC Chair, and General Chair of over 80 international conferences. He received nine Best Paper Awards from the IEEE Globecom 2010, the IEEE ICCT 2011, ITST 2012, the IEEE VTC 2013, IWCMC 2015, IWCMC 2016, the IEEE/CIC ICCC 2016, and the WPMC 2016. He is recognized as a Web of Science 2017 Highly Cited Researcher.
\end{IEEEbiography}

\begin{IEEEbiography}[{\includegraphics[width=1.1in,height=1.3in,clip,keepaspectratio] {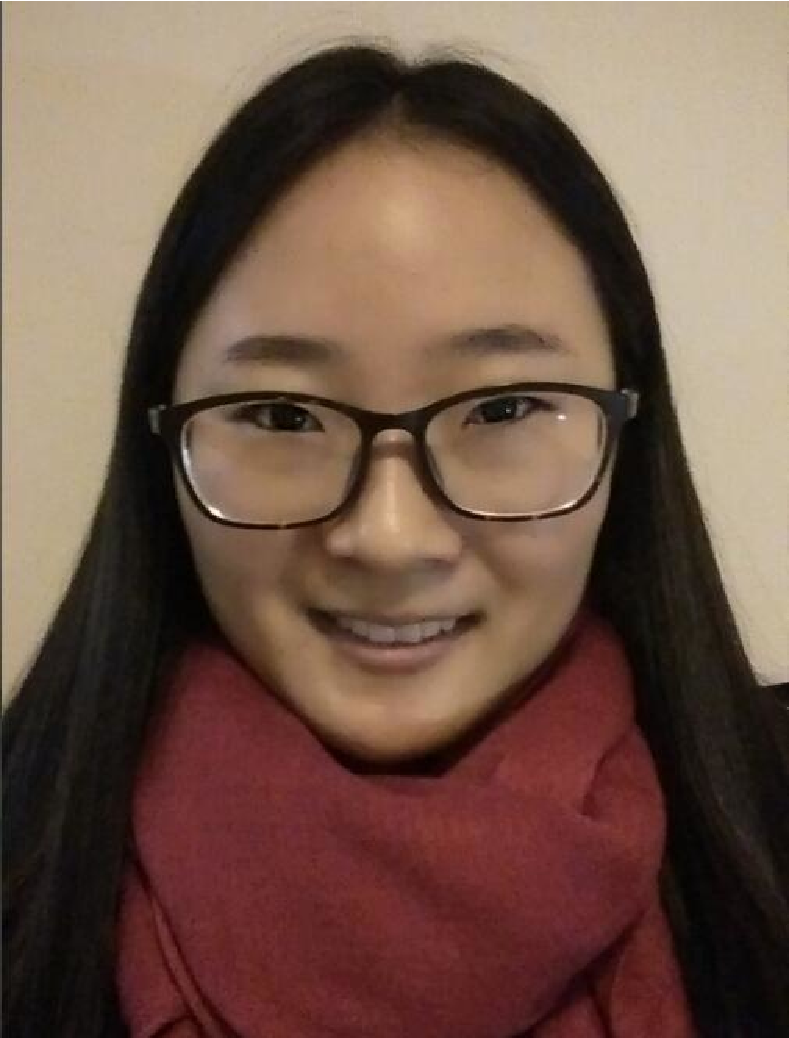}}]
{Rui Feng}received the B.Sc. degree in communication engineering and the M.Eng. degree in signal and information processing from Yantai University, Yantai, China, in 2011 and 2014, respectively.

She is currently pursuing the Ph.D. degree with the School of Information Science and Engineering, Shandong University, Shandong, China. Her current research interests include channel parameter estimation, millimeter wave and massive multiple-input multiple-output channel measurements and modeling.
\end{IEEEbiography}

\begin{IEEEbiography}[{\includegraphics[width=1.1in,height=1.3in,clip,keepaspectratio]{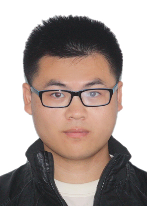}}]
{Jie Huang}received the BSc degree in Information Engineering from Xidian University, China, in 2013.

He is currently pursuing the PhD degree in Communication and Information Systems with Shandong University, China. His current research interests include millimeter wave and massive MIMO channel measurements, parameter estimation, channel modeling, wireless big data, and 5G wireless communications.
\end{IEEEbiography}
\vspace{-3cm}
\begin{IEEEbiography}[{\includegraphics[width=1.1in,height=1.3in,clip,keepaspectratio]{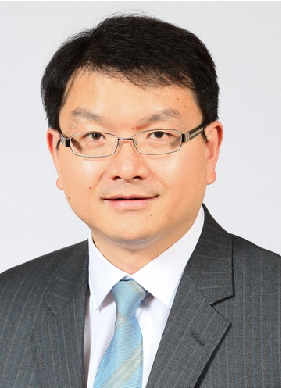}}]
{Yang Yang} (S'99-M'02-SM'10-F'18) is currently a professor with Shanghai Institute of Microsystem and Information Technology (SIMIT), Chinese Academy of Sciences, serving as the Director of CAS Key Laboratory of Wireless Sensor Network and Communication, and the Director of Shanghai Research Center for Wireless Communications (WiCO). He is also a Distinguished Adjunct Professor with the School of Information Science and Technology, ShanghaiTech University. Prior to that, he has held faculty positions at The Chinese University of Hong Kong, Brunel University, and University College London (UCL), UK.

Yang is a member of the Chief Technical Committee of the National Science and Technology Major Project ``New Generation Mobile Wireless Broadband Communication Networks" (2008-2020), which is funded by the Ministry of Industry and Information Technology (MIIT) of China. In addition, he is on the Chief Technical Committee for the National 863 Hi-Tech R\&D Program ``5G System R\&D Major Projects", which is funded by the Ministry of Science and Technology (MOST) of China. Since January 2017, he has been serving the OpenFog Consortium as the Director for Greater China Region.

Yang's current research interests include wireless sensor networks, Internet of Things, Fog computing, Open 5G, and advanced wireless testbeds. He has published more than 150 papers and filed over 80 technical patents in wireless communications. He is a Fellow of the IEEE.
\end{IEEEbiography}
\vspace{-3cm}
\begin{IEEEbiography}[{\includegraphics[width=1.1in,height=1.3in,clip,keepaspectratio]{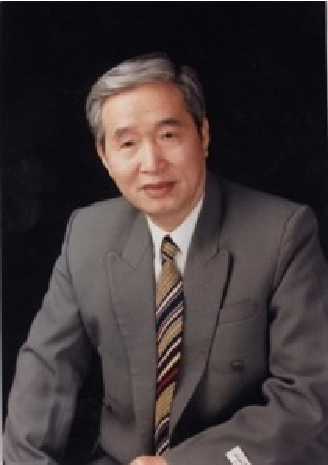}}]
{Minggao Zhang} received the BSc degree in mathematics from Wuhan University, China, in 1962.

He is currently a Distinguished Professor with the School of Information Science and Engineering, Shandong University, director of Academic Committee of China Rainbow Project Collaborative Innovation Center, director of Academic Committee of Shandong Provincial Key Lab of Wireless Communication Technologies, and a senior engineer of No. 22 Research Institute of China Electronics Technology Corporation (CETC). He has been an academician of Chinese Academy of Engineering since 1999 and is currently a Fellow of China Institute of Communications (CIC). He was a group leader of the Radio Transmission Research Group of ITU-R.

Minggao Zhang has been engaged in the research of radio propagation for decades. Many of his proposals have been adopted by international standardization organizations, including CCIR P.617-1, ITU-R P.680-3, ITU-R P.531-5, ITU-R P.529-2, and ITU-R P.676-3. In addition, he has received seven national and ministerial-level Science and Technology Progress Awards in China.
\end{IEEEbiography}





\end{document}